\pdfoutput=1
\documentclass[prx,amsmath,amssymb,superscriptaddress,floatfix,notitlepage,nofootinbib]{revtex4-1}

\usepackage{amsfonts}
\usepackage{graphicx}
\usepackage{graphicx}%Include figure files
\usepackage{dcolumn}%Align table columns on decimal point
\usepackage{bm}% bold math
\usepackage{hyperref}

\begin{document}

	\title{First-principles quantum transport modeling of spin-transfer and spin-orbit torques in magnetic multilayers}
	
	\author{Branislav K. Nikoli\'{c}}
	\email{bnikolic@udel.edu}
	\affiliation{Department of Physics and Astronomy, University of Delaware, Newark, DE 19716, USA}
	\author{Kapildeb Dolui}
	\affiliation{Department of Physics and Astronomy, University of Delaware, Newark, DE 19716, USA}
	\author{Marko Petrovi\'{c}}
	\affiliation{Department of Mathematical Sciences, University of Delaware, Newark,  DE 19716, USA}
	\author{Petr Plech\'a\v{c}}
	\affiliation{Department of Mathematical Sciences, University of Delaware, Newark,  DE 19716, USA}
	\author{Troels Markussen}
	\affiliation{Synopsys QuantumWise, Fruebjergvej 3, DK-2100 Copenhagen, Denmark}
	\author{Kurt Stokbro}
	\affiliation{Synopsys QuantumWise, Fruebjergvej 3, DK-2100 Copenhagen, Denmark}

\begin{abstract}
We review a unified approach for computing: ({\em i}) spin-transfer torque in magnetic trilayers like spin-valves and magnetic tunnel junction, where injected charge current flows perpendicularly to interfaces; and ({\em ii}) spin-orbit torque in magnetic bilayers of the type  ferromagnet/spin-orbit-coupled-material, where injected charge current flows parallel to the interface. The experimentally explored and technologically relevant spin-orbit-coupled-materials include $5d$ heavy metals, topological insulators, Weyl semimetals and transition metal dichalcogenides. Our approach requires to construct the torque operator for a given Hamiltonian of the device and the steady-state nonequilibrium density matrix, where the latter is expressed in terms of the nonequilibrium Green's functions and split into three contributions. Tracing these contributions with the torque operator automatically yields field-like and damping-like components of spin-transfer torque or spin-orbit torque  vector, which is particularly advantageous for spin-orbit torque where the direction of these components depends on the unknown-in-advance orientation of the current-driven nonequilibrium spin density in the presence of spin-orbit coupling.  We provide illustrative examples by computing spin-transfer torque in a one-dimensional toy model of a magnetic tunnel junction and realistic Co/Cu/Co spin-valve, both of which are described by  first-principles Hamiltonians obtained from noncollinear density functional theory calculations;  as well as spin-orbit torque in a ferromagnetic layer described by a tight-binding Hamiltonian which includes spin-orbit proximity effect within ferromagnetic monolayers assumed to be generated by the adjacent monolayer transition metal dichalcogenide. In addition, we show how spin-orbit proximity effect, quantified by computing (via first-principles retarded Green's function) spectral functions and spin textures  on monolayers of realistic ferromagnetic material like Co in contact with heavy metal or monolayer transition metal dichalcogenide, can be tailored to enhance the magnitude of spin-orbit torque. We also quantify errors made in the calculation of spin-transfer torque when using Hamiltonian from collinear density functional theory, with rigidly rotated magnetic moments to create noncollinear magnetization configurations, instead of proper (but computationally more expensive) self-consistent Hamiltonian obtained from noncollinear density functional theory.
\end{abstract}

\maketitle
	
\section{What is spin torque and why is it useful?}\label{sec:intro}

\begin{figure}
	\includegraphics[scale=0.4,angle=0]{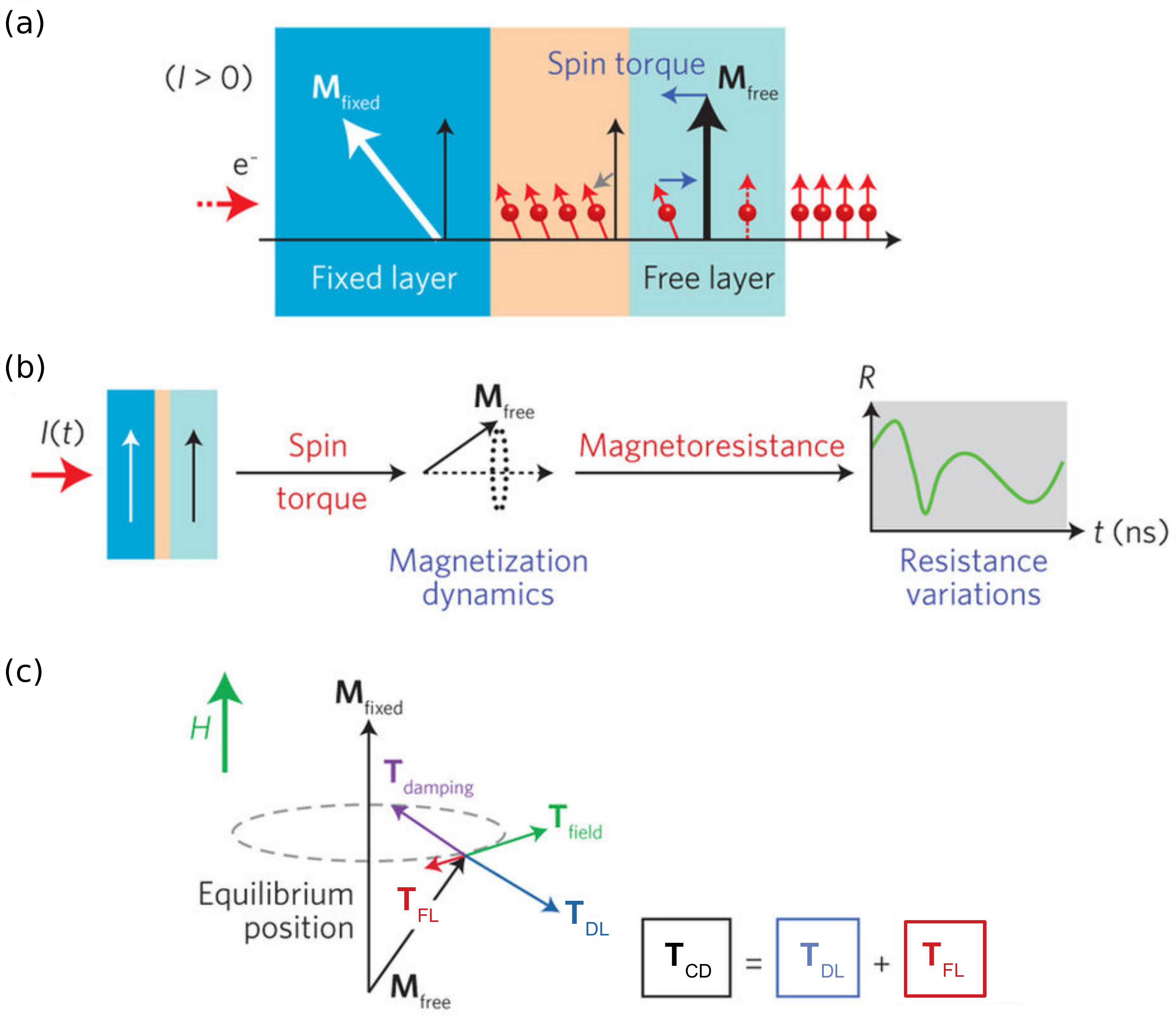}
	\caption{Spin-transfer torque basics: (a) in a ferromagnet/non-magnetic-spacer/ferromagnet setup, with noncollinear magnetizations $\mathbf{M}_\mathrm{fixed}$ of the fixed FM layer and $\mathbf{M}_\mathrm{free}$ of the free FM layer,  the transverse spin component of the conduction electrons (red) polarized in the direction of $\mathbf{M}_\mathrm{fixed}$ is absorbed as they pass through the free layer, thereby generating a torque on $\mathbf{M}_\mathrm{free}$; (b) device applications of STT in (a) are based on torque-induced magnetization dynamics that is converted into resistance variations via the magnetoresistive effects; (c) torques on $\mathbf{M}_\mathrm{free}$ where field-like component of current-driven STT, $\mathbf{T}_\mathrm{FL}$, is orthogonal to the plane spanned by $\mathbf{M}_\mathrm{fixed}$ and $\mathbf{M}_\mathrm{free}$ and competes with the effective-field torque  $\mathbf{T}_\mathrm{field}$ (present also in equilibrium), while damping-like component of current-driven STT, $\mathbf{T}_\mathrm{DL}$, is parallel or antiparallel (depending on the current direction) to Gilbert damping torque $\mathbf{T}_\mathrm{damping}$ (present also in equilibrium). The illustration in (c) assumes particular case where  $\mathbf{M}_\mathrm{fixed}$  and the effective magnetic field are aligned. Adapted from Ref.~\cite{Locatelli2014}.}
	\label{fig:fig0}
\end{figure}

\begin{figure}
	\includegraphics[scale=0.5,angle=0]{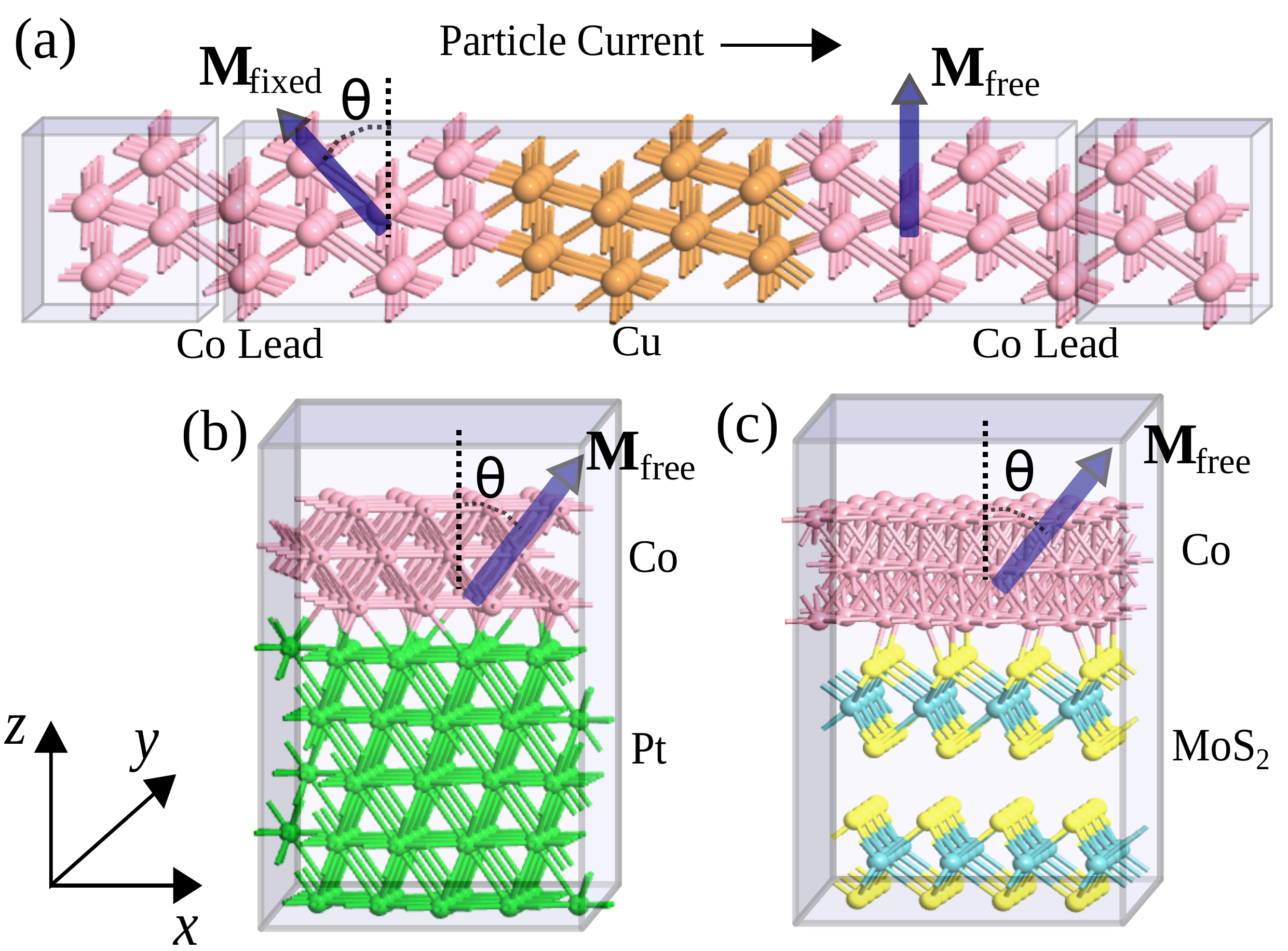}
	\caption{Schematic view of: (a) FM/NM/FM trilayer for calculations of STT in spin-valves; (b) FM/HM bilayer for calculations of SOT in the presence of the spin Hall current along the $z$-axis generated by the HM layer; and (c) FM/monolayer-TMD for calculations of SOT in the absence of any spin Hall current. The semi-infinite FM layers in (a) are chosen as Co(0001), and the spacer in between consists of $l$ ($l=4$ in the illustration) monolayers of Cu(111). The trilayer in (a) is assumed to be infinite in the transverse directions to the current flow, so that the depicted supercell is periodically repeated within the $yz$-plane. The bilayer in (b) consists of Co(0001) and Pt(111), and in (c) it consists of Co(0001) and monolayer MoS$_2$. The bilayers in (b) and (c) are assumed to be infinite in the $xy$-plane, but of finite thickness along the $z$-axis.  Small bias voltage $V_b$ is applied to inject electrons along the positive $x$-axis, so that particle current is perpendicular to interfaces in (a) and parallel to the interface in (b) and (c).}
	\label{fig:fig1}
\end{figure}

The spin-transfer torque (STT)~\cite{Ralph2008,Locatelli2014,Slonczewski1996,Berger1996} is a phenomenon in which a spin current of sufficiently large density ($\sim 10^7$ A/cm$^2$) injected into a ferromagnetic metal (FM)\footnote{For easy navigation, we provide a list of abbreviations used throughout the Chapter: 1D---one-dimensional; 2D---two-dimensional; 3D---three-dimensional; BZ---Brillouin zone; CD---current-driven; DFT---density functional theory; DL---damping-like; FL---field-like; FM---ferromagnetic metal; GF---Green's function; HM---heavy-metal; I---insulator; KS---Kohn-Sham; LCAO---linear combination of atomic orbitals; LLG---Landau-Lifshitz-Gilbert;  ML---monolayer; MLWF---maximally localized Wannier functions; MRAM---magnetic random access memory; MTJ---magnetic tunnel junction; ncDFT---noncollinear DFT; NEGF---nonequilibrium Green's function; NM---normal metal; PBE---Perdew-Burke-Ernzerhof; scf---self-consistent field; SHE---spin Hall effect; SOC---spin-orbit coupling; SOT---spin-orbit torque; STT---spin-transfer torque; TBH---tight-binding Hamiltonian; TI---topological insulator; TMD---transition metal dichalcogenide; WSM---Weyl semimetal; XC---exchange-correlation.} either switches its magnetization from one static configuration to another or generates a dynamical situation with steady-state precessing magnetization~\cite{Locatelli2014}. The origin of STT is the absorption of the itinerant flow of spin angular momentum component normal to the magnetization direction. Figure~\ref{fig:fig0}(a) illustrates a setup with two noncollinear magnetizations which generates STT. This setup can be realized as FM/NM/FM (NM-normal meal) spin-valve, exemplified by Co/Cu/Co trilayer in Fig.~\ref{fig:fig1}(a) and  employed in early experiments~\cite{Tsoi1998,Myers1999,Katine2000}; or FM/I/FM (I-insulator) magnetic tunnel junctions (MTJs), exemplified by Fe/MgO/Fe trilayer and employed in later experiments~\cite{Sankey2008,Kubota2008,Wang2011} and recent  applications~\cite{Locatelli2014,Kent2015}. In such magnetic multilayers, injected unpolarized charge current passes through the first thin FM layer to become spin-polarized in the direction of its fixed magnetization along the unit vector $\mathbf{M}_\mathrm{fixed}$, and it is directed into the second thin FM layer with magnetization along the unit vector $\mathbf{M}_\mathrm{free}$ where transverse (to $\mathbf{M}_\mathrm{free}$) component of flowing spins is absorbed. The STT-induced magnetization dynamics is converted into resistance variations via the magnetoresistive effect, as illustrated in Fig.~\ref{fig:fig0}(b), which is much larger in MTJ than in spin-valves. The rich nonequilibrium physics arising in the interplay of spin currents carried by fast conduction electrons, described quantum mechanically, and slow collective magnetization dynamics, described by the Landau-Lifshitz-Gilbert (LLG) equation which models magnetization as a classical vector subject to thermal fluctuations~\cite{Berkov2008,Evans2014,Petrovic2018}, is also of great fundamental interest. Note that at cryogenic temperatures, where thermal fluctuations are suppressed, quantum-mechanical effects in STT-driven magnetization dynamics can also be observed~\cite{Zholud2017,Mahfouzi2017,Mahfouzi2017a}.  

Another setup exhibiting current-induced magnetization dynamics is illustrated in Figs.~\ref{fig:fig1}(b) and ~\ref{fig:fig1}(c). It utilizes  a single FM layer, so that the role of polarizing FM layer with $\mathbf{M}_\mathrm{fixed}$ in Figs.~\ref{fig:fig0}(a) and ~\ref{fig:fig1}(a) is taken over by strong spin-orbit coupling (SOC) introduced by heavy metals (HMs)~\cite{Miron2011,Liu2012c} (such as $5d$ metals Pt, W and Ta) as in Fig.~\ref{fig:fig1}(b), topological insulators (TIs)~\cite{Mellnik2014,Fan2014a,Han2017,Wang2017}, Weyl semimetals (WSMs)~\cite{MacNeill2017,MacNeill2017a}, and even atomically thin transition metal dichalcogenides (TMDs)~\cite{Sklenar2016,Shao2016,Guimaraes2018,Lv2018}. The TMDs are compounds of the  type MX$_2$ (M = Mo, W, Nb; X = S, Se, Te) where one layer of M atoms is sandwiched between two layers of X atoms, as illustrated by monolayer MoS$_2$ in Fig.~\ref{fig:fig1}(c). The SOC is capable of converting charge into spin currents~\cite{Vignale2010,Sinova2015,Soumyanarayanan2016}, so that their absorption by the FM layer in Figs.~\ref{fig:fig1}(b) and ~\ref{fig:fig1}(c) leads to the so-called spin-orbit torque (SOT)~\cite{Manchon2018} on its free magnetization $\mathbf{M}_\mathrm{free}$. 

The current-driven (CD) STT and SOT vectors are analyzed by decomposing them into two contributions, \mbox{$\mathbf{T}_\mathrm{CD}=\mathbf{T}_\mathrm{DL} + \mathbf{T}_\mathrm{FL}$}, commonly termed~\cite{Ralph2008,Manchon2018} damping-like (DL) and field-like (FL) torque based on how they enter into the LLG equation describing the classical dynamics of magnetization. As illustrated in Fig.~\ref{fig:fig0}(c), these two torque components provide two different handles to manipulate the dynamics of $\mathbf{M}_\mathrm{free}$. In the absence of current, displacing $\mathbf{M}_\mathrm{free}$ out of its equilibrium position leads to the effective-field torque $\mathbf{T}_\mathrm{field}$   which  drives $\mathbf{M}_\mathrm{free}$ into precession around the effective magnetic field, while Gilbert damping $\mathbf{T}_\mathrm{damping}$ acts to bring it back to its equilibrium position. Under nonequilibrium conditions, brought by injecting steady-state or pulse current~\cite{Baumgartner2017}, $\mathbf{T}_\mathrm{DL}$ acts opposite to $\mathbf{T}_\mathrm{damping}$ for ``fixed-to-free'' current direction and it enhances $\mathbf{T}_\mathrm{damping}$ for ``free-to-fixed'' current directions in Fig.~\ref{fig:fig0}(a). Thus, the former (latter) acts as antidamping (overdamping) torque trying to bring $\mathbf{M}_\mathrm{free}$ antiparallel (parallel) to $\mathbf{M}_\mathrm{fixed}$ (note that at cryogenic temperatures one finds apparently only antidamping action of  $\mathbf{T}_\mathrm{DL}$ for both current directions~\cite{Zholud2017}). The $\mathbf{T}_\mathrm{FL}$ component induces magnetization precession and modifies the energy landscape seen by $\mathbf{M}_\mathrm{free}$. Although $|\mathbf{T}_\mathrm{FL}|$ is minuscule in metallic spin-valves~\cite{Wang2008b}, it can reach 30--40\% of $|\mathbf{T}_\mathrm{DL}|$ in MTJs~\cite{Sankey2008,Kubota2008}, and it can become several times larger than $|\mathbf{T}_\mathrm{DL}|$ in FM/HM bilayers~\cite{Kim2013,Yoon2017}. Thus, $\mathbf{T}_\mathrm{FL}$ component of SOT can play a crucial role~\cite{Baumgartner2017,Yoon2017} in triggering the reversal process of $\mathbf{M}_\mathrm{free}$ and in enhancing the switching efficiency. In concerted action with $\mathbf{T}_\mathrm{DL}$ and possible other effects brought by interfacial SOC, such as the Dzyaloshinskii-Moriya interaction~\cite{Perez2014}, this can also lead to complex inhomogeneous magnetization switching patterns observed in SOT-operated devices~\cite{Baumgartner2017,Yoon2017,Perez2014}.

By adjusting the ratio $|\mathbf{T}_\mathrm{DL}|/|\mathbf{T}_\mathrm{FL}|$~\cite{Timopheev2015} via tailoring of material properties and device shape, as well as by tuning the amplitude and duration of the injected pulse current~\cite{Baumgartner2017},  both STT- and SOT-operated devices can implement variety of functionalities, such as: nonvolatile magnetic random access memories (MRAM) of almost unlimited endurance; microwave oscillators; microwave detectors; spin-wave emitters; memristors; and artificial neural networks~\cite{Locatelli2014,Kent2015,Borders2017}. The key goal in all such  applications is to actively manipulate magnetization dynamics,  without the need for external magnetic fields that are incompatible with downscaling of the device size, while using the smallest possible current (e.g., writing currents $\lesssim 20$ $\mu$A would enable multigigabit MRAM~\cite{Kent2015}) and energy consumption. For example, recent experiments~\cite{Wang2017} have demonstrated current-driven magnetization switching at room temperature in FM/TI bilayers using current density $\sim 10^5$ A/cm$^2$, which is two orders of magnitude smaller than for STT-induced magnetization switching in MTJs or one to two orders of magnitude smaller than for SOT-induced magnetization switching in FM/HM bilayers. The SOT-MRAM is expected to  be less affected by damping, which offers flexibility for choosing the FM layer, while it eliminates insulating barrier in the writing process and its possible dielectric breakdown in STT-MRAM based on MTJs~\cite{Kent2015}. Also, symmetric switching profile of SOT-MRAM evades the asymmetric switching issues in STT-MRAM otherwise requiring additional device/circuit engineering. On the other hand, SOT-MRAM has a disadvantage of being a three-terminal device.

\section{How to model spin torque using nonequilibrium density matrix combined with density functional theory calculations}\label{sec:dm}

The absorption of the component of flowing spin angular momentum that is transverse to $\mathbf{M}_\mathrm{free}$, as illustrated in Fig.~\ref{fig:fig0}(a), occurs~\cite{Stiles2002,Wang2008b} within a few ferromagnetic monolayers (MLs) near NM/FM or I/FM interface. Since the thickness of this interfacial region is typically shorter~\cite{Wang2008b} than any charge or spin dephasing length that would make electronic transport semiclassical, STT {\em requires} quantum transport modeling~\cite{Brataas2006}. The essence of STT can be understood using simple one-dimensional (1D) models solved by matching spin-dependent wave function across the junction, akin to elementary quantum mechanics problems of transmission and reflection through a barrier, as provided in Refs.~\cite{Ralph2008,Manchon2008a,Xiao2008a}. However, to describe details of experiments, such as bias voltage dependence of STT in  MTJs~\cite{Kubota2008,Sankey2008} or complex angular dependence of SOT in FM/HM bilayers~\cite{Garello2013}, more involved calculations are needed employing tight-binding or first-principles Hamiltonian as an input. For example, simplistic tight-binding Hamiltonians (TBHs) with single orbital per site have been coupled~\cite{Theodonis2006} to nonequilibrium Green's function (NEGF) formalism~\cite{Stefanucci2013} to compute SOT in FM/HM bilayers~\cite{Kalitsov2017}, or bias voltage dependence of DL and FL components of STT in MTJs which can describe some features of the experiments by adjusting the tight-binding parameters~\cite{Kubota2008}. 

\begin{figure}
	\includegraphics[scale=0.5,angle=0]{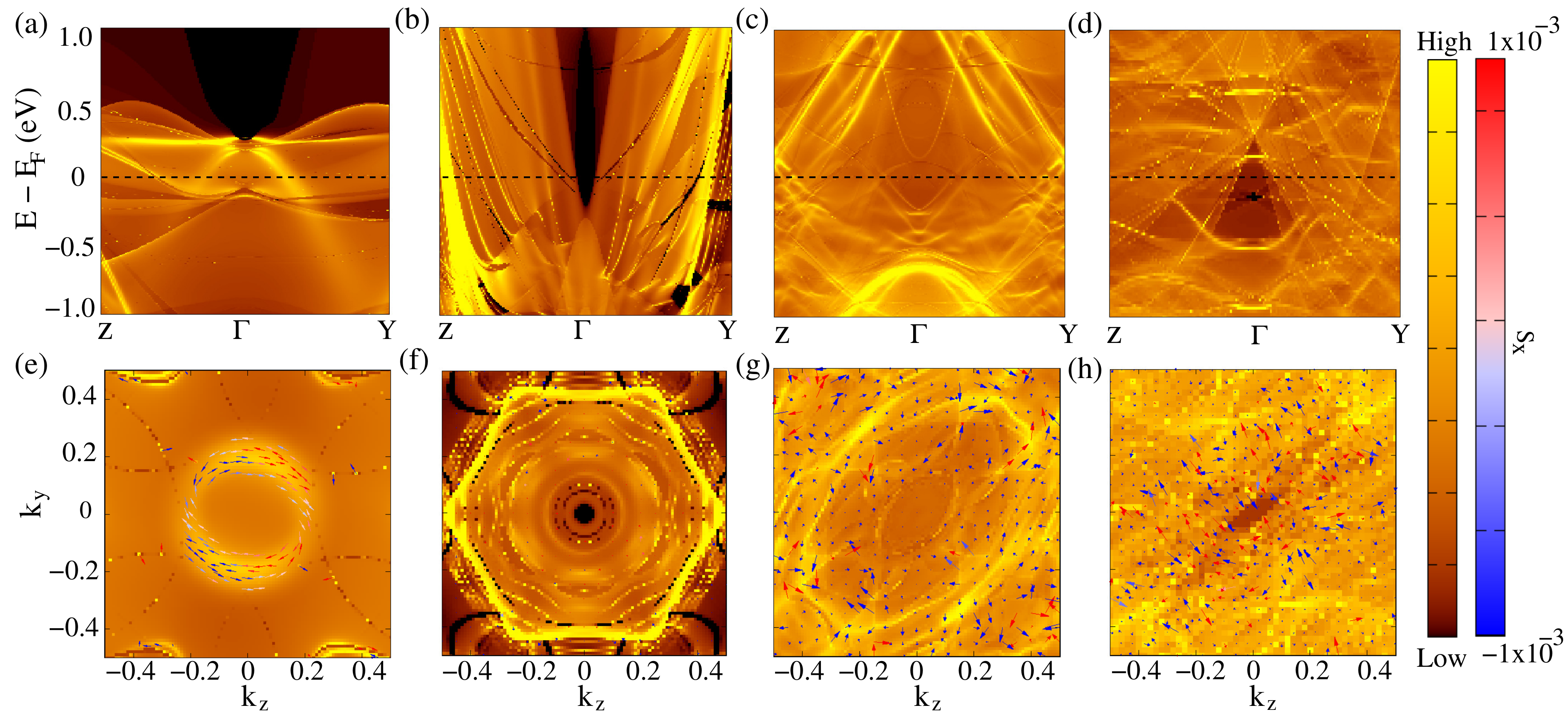}
	\caption{Spectral function $A(E; \mathbf{k}_\parallel, x \in \mathrm{ML \ of  \ Co})$, defined in Eq.~\eqref{eq:spectral}, plotted along high symmetry $k$-path $Z$-$\Gamma$-$Y$ at: (a) the surface of semi-infinite Co(0001) in contact with vacuum; (b) ML of semi-infinite Co(0001) in contact with 9 MLs of Cu(111) within Co/Cu(9 ML)/Co spin-valve;  (c) ML of semi-infinite Co(0001) in contact with 5 MLs of Pt(111); and (d) ML of semi-infinite Co(0001) in contact with 1 ML of MoS$_2$. Panels (e)--(h) plot constant energy contours of \mbox{$A(E-E_F =0;\mathbf{k}_\parallel, x \in \mathrm{ML \ of  \ Co})$} and the corresponding spin textures where the out-of-plane $S_x$ component of spin is indicated in color (red for positive and blue for negative). The magnetization of Co is along the $x$-axis which is perpendicular to the ML of Co over which the spectral functions and spin textures are scanned. The units for $k_y$ and $k_z$ are $2\pi/a$ and $2\pi/b$ where $a$ and $b$ are the lattice constants along the $y$- and the $z$-axis, respectively. The horizontal dashed black line in panels (a)--(d) denotes the position of the Fermi energy $E_F$. Panels (a) and (e) are adapted from Ref.~\cite{Marmolejo-Tejada2017}, and panels (c) and (g) are adapted from Ref.~\cite{Dolui2017}.}
	\label{fig:fig2}
\end{figure}

However, not all features of STT experiments on MTJs~\cite{Wang2011} can be captured by such NEGF+TBH approach. Furthermore, due to spin-orbit proximity effect, driven by hybridization of wave functions from FM and HM layers~\cite{Dolui2017} or FM and metallic surfaces of three-dimensional (3D) TIs~\cite{Marmolejo-Tejada2017}, simplistic Hamiltonians like the Rashba ferromagnetic  model~\cite{Manchon2008,Haney2013,Lee2015,Li2015,Pesin2012a,Ado2017,Kalitsov2017} or the gapped Dirac model~\cite{Ndiaye2017} are highly inadequate to describe realistic bilayers employed in SOT experiments. This is emphasized by Fig.~\ref{fig:fig2} which shows how spectral function and spin texture on the surface of semi-infinite Co layer can change dramatically as we change the adjacent layer. For example, nonzero in-plane spin texture on the surface of semi-infinite Co layer in contact with vacuum is found in Fig.~\ref{fig:fig2}(e), despite Co magnetization being perpendicular to that surface. This is the consequence of the Rashba SOC enabled by inversion symmetry breaking~\cite{Chantis2007} where an electrostatic potential gradient can be created by the charge distribution at the metal/vacuum interface to  confine wave functions into a Rashba spin-split quasi two-dimensional (2D) electron gas~\cite{Bahramy2012}. The surface of semi-infinite Co within layer embedded into Co/Cu(9 ML)/Co junction illustrated in Fig.~\ref{fig:fig1}(a) does not have any in-plane spin texture in Fig.~\ref{fig:fig2}(f) since the structure is inversion symmetric, but its spectral function in Fig.~\ref{fig:fig2}(b) is quite different from the one on the surface of an isolated semi-infinite Co layer in Fig.~\ref{fig:fig2}(a). Bringing semi-infinite Co layer in contact with 5 MLs of Pt or 1 ML of MoS$_2$ transforms its spectral function from Fig.~\ref{fig:fig2}(a) to the ones in Figs.~\ref{fig:fig2}(c) and ~\ref{fig:fig2}(d), respectively, while inducing the corresponding spin textures in Figs.~\ref{fig:fig2}(g) and ~\ref{fig:fig2}(h) due to spin-orbit proximity effect signified by the ``leakage'' of SOC from HM or TMD layer into the FM layer.  

Note that the spectral function and spin texture at the Co/Pt interface are quite different from those of the ferromagnetic Rashba Hamiltonian in 2D often employed~\cite{Manchon2008,Haney2013,Lee2015,Li2015,Pesin2012a,Ado2017,Kalitsov2017} in the calculations of SOT as the putative simplistic description of the FM/HM interface. When charge current flows within FM monolayer hosting spin textures---such as the ones displayed in Figs.~\ref{fig:fig2}(e),  ~\ref{fig:fig2}(g) and~\ref{fig:fig2}(h)---more forward-going electron states will be occupied and less the backward going ones which due to spin-momentum locking leads to nonequilibrium spin density~\cite{Edelstein1990,Aronov1989} as one of the principal mechanisms behind SOT~\cite{Manchon2018}. The direction of the nonequilibrium spin density is easily identified in the case of simple spin textures, such as the one in Fig.~\ref{fig:fig2}(e) or those associated with simplistic models~\cite{Pesin2012} like the Rashba Hamiltonian or the Dirac Hamiltonian discussed in Sec.~\ref{sec:sot}. Conversely, for complex spin textures within heterostructures of realistic materials, as exemplified by those in  Figs.~\ref{fig:fig2}(g) and ~\ref{fig:fig2}(h), one needs first-principles coupled with electronic transport calculations~\cite{Chang2015,Johansson2018}.

Thus, capturing properties of realistic junctions illustrated by Fig.~\ref{fig:fig2} requires first-principles Hamiltonian as offered by the density functional theory (DFT). In the linear-response regime,  appropriate for spin-valves or SOT-operated bilayers in Fig.~\ref{fig:fig1}, one can also employ first-principles derived TBH as offered by transforming the DFT Hamiltonian to a basis of orthogonal  maximally localized Wannier functions (MLWFs) in a selected energy window around the Fermi energy $E_F$. This procedure retains faithfully the overlap matrix elements and their phases, orbital character of the bands and the accuracy of the original DFT calculations~\cite{Marzari2012}. While Wannier TBH has been used to describe infinite-FM-on-infinite-HM bilayers~\cite{Freimuth2014,Mahfouzi2018}, its accuracy can be compromised by complicated band entanglement in hybridized metallic systems~\cite{Marzari2012}. It is also cumbersome to construct Wannier TBH for junctions in other geometries, like the spin-valve in Fig.~\ref{fig:fig1}(a) or when FM/HM bilayer is attached to leads made of different NM material. In such cases, one needs to perform multiple calculations~\cite{Shelley2011,Thygesen2005a} (such as on periodic leads,  supercell composed of the central region of interest  attached to buffer layers of the lead material on both sides, etc.) where one can encounter different MLWFs for two similar but nonidentical systems~\cite{Thygesen2005a}; nonorthogonal MLWFs belonging to two different regions~\cite{Thygesen2005a}; and Fermi energies of distinct calculations that have to be aligned~\cite{Shelley2011}. Also, to compute current or STT in MTJs at finite bias voltage one needs to recalculate Hamiltonian in order to take into account self-consistent charge redistribution and the corresponding electrostatic potential in the presence of current flow. Otherwise, without computing them across the device the current-voltage characteristics violates~\cite{Christen1996,Hernandez2013} gauge invariance, i.e., invariance with respect to the global shift of electric potential by a constant, $V \rightarrow V + V_0$. 

The noncollinear DFT (ncDFT)~\cite{Capelle2001,Eich2013a,Eich2013,Bulik2013} coupled to nonequilibrium density matrix~\cite{Stefanucci2013} offers an algorithm to compute spin torque in arbitrary device geometry at vanishing or finite bias voltage. The single-particle spin-dependent Kohn-Sham (KS) Hamiltonian in ncDFT takes the form
\begin{equation}\label{eq:ncdft}
\hat{H}_\mathrm{KS} = -\frac{\hbar^2\nabla^2}{2m}  + V_\mathrm{H}({\bf r}) + V_{\rm XC}({\bf r}) + V_{\rm ext}({\bf r}) - {\bm \sigma} \cdot \mathbf{B}_\mathrm{XC}(\mathbf{r}),
\end{equation}
where $V_H({\bf r})$, $V_{\rm ext}({\bf r})$ and $V_\mathrm{XC}({\bf r})=E_\mathrm{XC}[n(\mathbf{r}),\mathbf{m}(\mathbf{r})]/\delta n(\mathbf{r})$ are the Hartree, external and exchange-correlation (XC) potentials, respectively, and ${\bm \sigma}=(\hat{\sigma}_x,\hat{\sigma}_y,\hat{\sigma}_z)$ is the vector of the Pauli matrices. The extension of DFT to the case of spin-polarized systems is formally derived in terms of total electron density $n(\mathbf{r})$ and vector  magnetization density $\mathbf{m}(\mathbf{r})$. In the collinear DFT, $\mathbf{m}(\mathbf{r})$ points in the same direction at all points in space, which is insufficient to study magnetic systems where the direction of the local magnetization is not constrained to a particular axis or systems with SOC. In ncDFT~\cite{Capelle2001},  XC functional $E_\mathrm{XC}[n(\mathbf{r}),\mathbf{m}(\mathbf{r})]$ depends on $\mathbf{m}(\mathbf{r})$ pointing in arbitrary direction. The XC magnetic field is then given by \mbox{$\mathbf{B}_\mathrm{XC}(\mathbf{r}) = \delta E_\mathrm{XC}[n(\mathbf{r}),\mathbf{m}(\mathbf{r})]/\delta \mathbf{m}(\mathbf{r})$}.

Once the Hamiltonian of the device is selected, it has to be passed into the formalism of nonequilibrium quantum statistical mechanics. Its concept is the density matrix ${\bm  \rho}$  of quantum many-particle system at finite temperature, or in the presence of external static or time-dependent fields which drive the system out of equilibrium. The knowledge of ${\bm  \rho}$ makes it possible to compute the expectation value of any observable
\begin{equation}\label{eq:expectation}
O=\mathrm{Tr}\, [{\bm \rho} \mathbf{O}],
\end{equation}
such as the charge density, charge current, spin current and spin density of interest to spin torque modeling. These requires to insert their operators (in some matrix representation) as $\mathbf{O}$ into Eq.~\eqref{eq:expectation}, where we use notation in which bold letters denote matrix representation of an operator in a chosen basis. For the KS Hamiltonian in ncDFT in Eq.~\eqref{eq:ncdft}, the torque operator is given by the time derivative of the electronic spin operator~\cite{Haney2007,Carva2009}
\begin{equation}\label{eq:torqueop}
\mathbf{T} = \frac{d{\bf S}}{dt} = \frac{1}{2i} [{\bm \sigma},\mathbf{H}_\mathrm{KS}] = {\bm \sigma} \times \mathbf{B}_\mathrm{XC},
\end{equation}
Its trace with ${\bm \rho}$ yields the spin torque vector while concurrently offering a microscopic picture~\cite{Haney2007} for the origin of torque---misalignment of the nonequilibrium spin density of current carrying quasiparticles with respect to the spins of electrons comprising the magnetic condensate responsible for nonzero $\mathbf{B}_\mathrm{XC}$. This causes local torque on individual atoms, which is summed by performing trace in Eq.~\eqref{eq:expectation} to find the net effect on the total magnetization $\mathbf{M}_\mathrm{free}$ of the free FM layer. Examples of how to evaluate such trace, while using $\mathbf{O} \mapsto \mathbf{T}$ in Eq.~\eqref{eq:torqueop} in different matrix representations,  are given as Eqs.~\eqref{eq:lcaotrace} and ~\eqref{eq:realspacetrace} in Sec.~\ref{sec:stt}.

In equilibrium, ${\bm \rho}_\mathrm{eq}$ is fixed by the Boltzmann-Gibbs prescription, such as ${\bm \rho}_\mathrm{eq} = \sum_n f(E) |\Psi_n \rangle \langle \Psi_n|$ in grand canonical ensemble describing electrons with the Fermi distribution function $f(E)$ due to contact with a  macroscopic reservoir at chemical potential $\mu$ and temperature $T$, where $E_n$ and $|\Psi_n\rangle$ are eigenenergies and eigenstates of the Hamiltonian, respectively. Out of equilibrium, the construction of ${\bm \rho}_\mathrm{neq}$ is complicated by the variety of possible driving fields and open nature of a driven quantum system. For example, the Kubo linear-response theory has been used to obtain ${\bm \rho}_\mathrm{neq}$ for small applied electric field in infinite-FM-on-infinite-HM bilayer geometry~\cite{Freimuth2014,Mahfouzi2018}. However, for arbitrary junction geometry and magnitude of the applied bias voltage $V_b$ or injected pulse current, the most advantageous is to employ the NEGF formalism~\cite{Stefanucci2013}. This requires to evaluate its two fundamental objects---the retarded GF, \mbox{$G^{\sigma\sigma'}_{\mathbf{nn}'}(t,t')=-i \Theta(t-t') \langle \{\hat{c}_{\mathbf{n}\sigma}(t) , \hat{c}^\dagger_{\mathbf{n}'\sigma'}(t')\}\rangle$},  and the lesser GF, \mbox{$G^{<,\sigma\sigma'}_{\mathbf{nn}'}(t,t')=i \langle \hat{c}^\dagger_{\mathbf{n}'\sigma'}(t') \hat{c}_{\mathbf{n} \sigma}(t)\rangle$}---describing the density of available quantum states and how electrons occupy those states, respectively. The operator $\hat{c}_{\mathbf{n}\sigma}^{\dagger}$ ($\hat{c}_{\mathbf{n}\sigma}$) creates (annihilates) electron with spin $\sigma$ at site $\mathbf{n}$ (another index would be required to label more than one orbital present at the site), and  $\langle \ldots \rangle$ denotes the nonequilibrium statistical average~\cite{Stefanucci2013}. 

In time-dependent situations, the nonequilibrium density matrix is given by~\cite{Petrovic2018,Stefanucci2013} 
\begin{equation}\label{eq:rhoneqt}
{\bm \rho}_\mathrm{neq}(t) = \mathbf{G}^<(t,t)/i.
\end{equation}
In stationary problems $\mathbf{G}$ and $\mathbf{G}^<$ depend only on the time difference $t-t^\prime$ and can, therefore, be Fourier transformed to depend on energy $E$ instead of $t-t^\prime$. The retarded GF in stationary situations is then given by 
\begin{equation}\label{eq:rgf}
\mathbf{G}(E) = \left[ E- \mathbf{H} - {\bm \Sigma}_L(E,V_b) - {\bm \Sigma}_R(E,V_b) \right]^{-1}, 
\end{equation}
assuming representation in the basis of orthogonal orbitals. In the case of nonorthogonal basis set $|\phi_n \rangle$, one should make a replacement $E \mapsto E\mathbf{S}$ where $\mathbf{S}$ is the overlap matrix composed of elements $\langle \phi_n|\phi_m\rangle$. The self-energies~\cite{Velev2004,Rungger2008} ${\bm \Sigma}_{L,R}(E,V_b)$ describe the semi-infinite leads which guarantee continuous energy spectrum of devices in Fig.~\ref{fig:fig1} required to reach the steady-state transport regime. The leads terminate at infinity into the left (L) and right (R) macroscopic reservoirs with different electrochemical potentials, $\mu_L-\mu_R=eV_L-eV_R=eV_b$. The usual assumption about the leads is that the applied bias voltage $V_b$ induces a rigid shift in their electronic structure\cite{Brandbyge2002}, so that ${\bm \Sigma}_{L,R}(E,V_b)  =  {\bm \Sigma}_{L,R}(E-eV_{L,R})$. 

In equilibrium or near equilibrium (i.e., in the linear-response transport regime at small $eV_b \ll E_F$), one needs $\mathbf{G}_0(E)$ obtained from Eq.~\eqref{eq:rgf} by setting $V_L=V_R=0$. The spectral functions shown in Fig.~\ref{fig:fig2}(a)--(d) can be computed at an arbitrary plane at position $x$ within the junction in Fig.~\ref{fig:fig1}(a) using $\mathbf{G}_0(E)$ 
\begin{equation}\label{eq:spectral}
A(E;\mathbf{k}_\parallel,x)=-\frac{1}{\pi}\mathrm{Im}\,[G_0(E;{\mathbf{k}_\parallel};x,x)],
\end{equation} 
where the diagonal matrix elements $G_0(E;{\mathbf{k}_\parallel};x,x)$ are obtained by transforming the retarded GF from a local orbital to a real-space representation. The spin textures in  Fig.~\ref{fig:fig2}(e)--(h) within the constant energy contours are computed from the spin-resolved spectral function. The equilibrium density matrix can also be expressed in terms of $\mathbf{G}_0(E)$
\begin{equation}\label{eq:rhoeq}
{\bm \rho}_\mathrm{eq} = -\frac{1}{\pi} \int\limits_{-\infty}^{+\infty} dE \, {\rm Im}\, \mathbf{G}_0(E) f(E),
\end{equation}
where $\mathrm{Im}\, \mathbf{O} = (\mathbf{O} - \mathbf{O}^\dagger)/2i$. 

The nonequilibrium density matrix is determined by the lesser GF
\begin{equation}\label{eq:rhoneq}
{\bm \rho}_\mathrm{neq} = \frac{1}{2\pi i} \int\limits_{-\infty}^{+\infty} dE\, \mathbf{G}^<(E).
\end{equation}
In general, if a quantity has nonzero expectation values in equilibrium, that one must be subtracted from the final result since it is {\em unobservable} in transport experiments. This is  exemplified by spin current density in time-reversal invariant systems~\cite{Nikolic2006}; spin density, diamagnetic circulating currents and circulating heat currents  in the presence of external magnetic field or spontaneous magnetization breaking time-reversal invariance; and FL component of STT~\cite{Theodonis2006}. Thus, we define the current-driven  part of the nonequilibrium density matrix as
\begin{equation}\label{eq:rhocurr}
{\bm \rho}_\mathrm{CD} = {\bm \rho}_\mathrm{neq} - {\bm \rho}_\mathrm{eq}.
\end{equation}
Although the NEGF formalism can include many-body interactions, such as electron-magnon scattering~\cite{Mahfouzi2014} that can affect STT~\cite{Zholud2017,Levy2006,Manchon2010} and SOT~\cite{Yasuda2017,Okuma2017}, here we focus on the usually considered and conceptually simpler elastic transport regime where the lesser GF of a two-terminal junction 
\begin{equation}\label{eq:lgf}
\mathbf{G}^<(E)=\mathbf{G}(E) \left[i f_L(E) {\bm \Gamma}_L(E) +  i f_R(E) {\bm \Gamma}_R(E) \right] \mathbf{G}^\dagger(E),
\end{equation}
is expressed solely in terms of the retarded GF, the level broadening matrices \mbox{${\bm \Gamma}_{L,R}(E)=i[{\bm \Sigma}_{L,R}(E) - {\bm \Sigma}_{L,R}^\dagger(E)]$} determining the escape rates of electrons into the semi-infinite leads and shifted Fermi functions \mbox{$f_{L,R}(E)=f(E-eV_{L,R})$}.

For purely computational purposes, the integration in Eq.~\eqref{eq:rhoneq} is typically separated (non-uniquely~\cite{Xie2016}) into the apparent ``equilibrium'' and current-driven ``nonequilibrium''  terms~\cite{Brandbyge2002,Sanvito2011}
\begin{equation}\label{eq:rho}
{\bm \rho}_\mathrm{neq}   =  -\frac{1}{\pi} \int\limits_{-\infty}^{+\infty} dE \, {\rm Im}\, \mathbf{G}(E) f(E-eV_R) +  \frac{1}{2 \pi} \int\limits_{-\infty}^{+\infty}dE \,  \mathbf{G}(E) \cdot {\bm \Gamma}_L(E-eV_L) \cdot   \mathbf{G}^\dagger(E) \left[ f_L(E) - f_R(E) \right].
\end{equation}
The first ``equilibrium'' term contains integrand which is analytic in the upper complex plane and can be computed via contour  integration~\cite{Brandbyge2002,Areshkin2010,Ozaki2007,Karrasch2010}, while the integrand in the second ``current-driven'' term is nonanalytic function in the entire complex energy plane so that its integration has to be performed directly along the real axis~\cite{Sanvito2011}  between the limits set by the window of nonzero values of \mbox{$f_L(E) - f_R(E)$}. Although the second  term in Eq.~\eqref{eq:rho} contains information about the bias voltage [through the difference \mbox{$f_L(E)-f_R(E)$}] and about the lead assumed to be injecting electrons into the device (through ${\bm \Gamma}_L$), it cannot~\cite{Xie2016,Mahfouzi2013} be used as the proper ${\bm \rho}_\mathrm{CD}$ defined in Eq.~\eqref{eq:rhocurr}. This is due to the fact that second term in Eq.~\eqref{eq:rhocurr}, expressed in terms of the retarded GF via Eq.~\eqref{eq:rhoeq}, does not cancel the gauge-noninvariant first term in Eq.~\eqref{eq:rho} which depends explicitly [through $f(E-eV_R)$] on the arbitrarily chosen reference potential $V_R$ and implicitly on the voltages applied to both reservoirs [through $\mathbf{G}(E)$]. Nevertheless, the second term in Eq.~\eqref{eq:rho}, written in the linear-response and zero-temperature limit,
\begin{equation} \label{eq:improper}
{\bm \rho}_\mathrm{CD} \stackrel{?}{=} \frac{eV_b}{2 \pi} \mathbf{G}_0(E_F) \cdot {\bm \Gamma}_L(E_F) \cdot \mathbf{G}_0^\dagger(E_F),
\end{equation}
has often been used in STT literature~\cite{Haney2007,Heiliger2008b} as the putative but {\em improper} (due to being gauge-noninvariant, which we mark by using ``?'' on the top of the equality sign) expression for ${\bm \rho}_{\rm CD}$. Its usage  leads to ambiguous (i.e., dependent on arbitrarily chosen $V_R$) nonequilibrium expectation values. 

\begin{figure}
	\includegraphics[scale=0.4,angle=0]{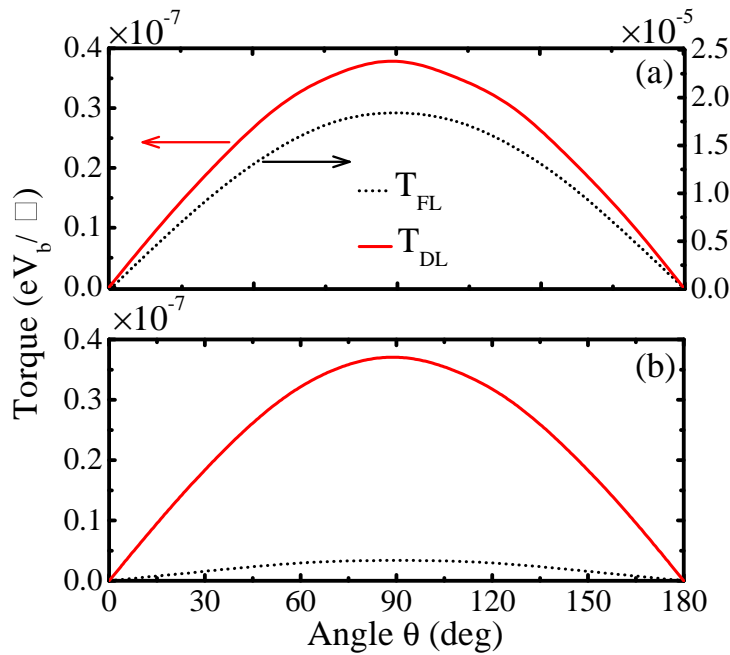}
	\caption{The angular dependence of the damping-like (i.e., parallel), $|\mathbf{T}_\mathrm{DL}|$, and field-like (i.e., perpendicular),  $|\mathbf{T}_\mathrm{FL}|$, components of the STT vector in Fig.~\ref{fig:fig0}(c) in an asymmetric~\cite{Oh2009} FM/I/FM$^\prime$ MTJ computed in the linear-response regime at zero-temperature using: (a) the proper gauge-invariant expression Eq.~\eqref{eq:proper} for ${\bm \rho}_\mathrm{CD}$; and (b) the improper gauge-noninvariant expression Eq.~\eqref{eq:improper} for ${\bm \rho}_\mathrm{CD}$. The FM/I/FM$^\prime$ trilayer in Fig.~\ref{fig:fig0}(a) is modeled by TBH~\cite{Theodonis2006} defined on an infinite cubic lattice with a single orbital per site and lattice spacing $a$. Its insulating barrier  has thickness 5 MLs with on-site potential $\varepsilon_n=6.0$ eV; the left FM layer is semi-infinite and the right FM$^\prime$ layer is 20 MLs thick. Both FM layers have the same exchange field $J=0.5$ eV. The symbol $\Box=a^2$ denotes unit interfacial area. Adapted from Ref.~\cite{Mahfouzi2013}.}
	\label{fig:fig3}
\end{figure}

The {\em proper} gauge invariant expressions was derived in Ref.~\cite{Mahfouzi2013}
\begin{equation}\label{eq:proper}
{\bm \rho}_{\rm CD}  =   \frac{eV_b}{2\pi} \mathbf{G}_0 (E_F) \cdot {\bm \Gamma}_L(E_F) \cdot \mathbf{G}_0^\dagger(E_F) - \frac{eV_R}{\pi} \mathrm{Im}\, \mathbf{G}_0(E_F) - \frac{1}{\pi} \int\limits_{-\infty}^{+\infty} \!\! dE \, \mathrm{Im} \left[ \mathbf{G}_0 \left(eU_\mathbf{n} - eV_L \frac{\partial {\bm \Sigma}_L}{\partial E} - eV_R \frac{\partial {\bm \Sigma}_R}{\partial E} \right) \mathbf{G}_0 \right] f(E),
\end{equation}
which we give here at zero-temperature so that it can be contrasted with Eq.~\eqref{eq:improper}. The second and third term in Eq.~\eqref{eq:proper}, whose purpose is to subtract any nonzero expectation value that exists in thermodynamic equilibrium, make it quite different from Eq.~\eqref{eq:improper} while requiring to include also electrostatic potential profile $U_\mathbf{n}$  across the active region of the device interpolating  between $V_L$ and $V_R$. For example, the second term in Eq.~\eqref{eq:proper} traced with an operator gives equilibrium expectation value governed by the states at $E_F$ which must be removed. The third term in Eq.~\eqref{eq:proper} ensures the gauge invariance of the nonequilibrium expectation values, while making the whole expression non-Fermi-surface property. The third term also renders the usage of  Eq.~\eqref{eq:proper} computationally demanding due to the requirement to perform integration from the bottom of the band up to $E_F$ together with sampling of $\mathbf{k}_\parallel$ points for the junctions in Fig.~\ref{fig:fig1}.

In Fig.~\ref{fig:fig3} we consider an example of a left-right {\em asymmetric} MTJ, FM/I/FM$^\prime$,  whose FM and FM$^\prime$ layers are assumed to be made of the same material but have different thicknesses. This setup allows us to demonstrate how application of improper ${\bm \rho}_{\rm CD}$ in Eq.~\eqref{eq:improper} yields linear-response  ${\bf T}_\mathrm{FL}  \propto V_b$ in Fig.~\ref{fig:fig3}(b) that is {\em incorrectly} an order of magnitude smaller than the correct result in Fig.~\ref{fig:fig3}(a). This is due to the fact that ${\bf T}_\mathrm{FL}$ in MTJs posses both the nonequilibrium CD contribution due to spin reorientation at interfaces, where net spin created at one interface is reflected at the second interface where it briefly precesses   in the exchange field of the free FM layer, and equilibrium contribution due to interlayer exchange coupling~\cite{Theodonis2006,Yang2010}. The ambiguity in Fig.~\ref{fig:fig3} arises when this equilibrium contribution is improperly subtracted, so that current-driven  ${\bf T}_\mathrm{FL}$ in Fig.~\ref{fig:fig3}(b) is contaminated by a portion of equilibrium contribution added to it when using improper ${\bm \rho}_\mathrm{CD}$ in Eq.~\eqref{eq:improper}. On the other hand, since ${\bf T}_\mathrm{DL}$ has a zero expectation value in equilibrium, both the proper and improper expressions for ${\bm \rho}_\mathrm{CD}$ give the same result in Fig.~\ref{fig:fig3}. 

Note that in the left-right {\em symmetric} MTJs, $\mathbf{T}_\mathrm{FL} \propto V_b$ vanishes. Since this is a rather general result which holds for both MTJs and spin-valves in the linear-response regime~\cite{Theodonis2006,Xiao2008a,Heiliger2008}, and it been confirmed in numerous  experiments~\cite{Wang2011,Oh2009}, one can use it as a validation test of the computational scheme. For example, the usage of improper ${\bm \rho}_\mathrm{CD}$ in Eq.~\eqref{eq:improper}, or the proper one in Eq.~\eqref{eq:proper} but with possible  software bug, would give nonzero ${\bf T}_\mathrm{FL} \neq 0$ in symmetric junctions at small applied $V_b$ which contradicts experiments~\cite{Wang2011,Oh2009}. In the particular case of symmetric junction one can actually employ a simpler expression~\cite{Mahfouzi2013,Stamenova2017} than Eq.~\eqref{eq:proper} which guarantees ${\bf T}_\mathrm{FL} \equiv 0$ 
\begin{equation}\label{eq:symmetricdm}
{\bm \rho}_\mathrm{CD} = \frac{eV_b}{4 \pi} \mathbf{G}_0(E_F) \cdot [{\bm \Gamma}_L(E_F) -  {\bm \Gamma}_R(E_F)]\cdot \mathbf{G}_0^\dagger(E_F),
\end{equation} 
obtained by assuming~\cite{Mahfouzi2013} the particular gauge $V_L=-V_b/2=-V_R$. Such special gauges and the corresponding Fermi surface expressions for ${\bm \rho}_\mathrm{CD} \propto V_b$ in the linear-response regime do exist also for asymmetric junctions, but one does not know them in advance except for the special case of symmetric junctions~\cite{Mahfouzi2013}.

\begin{figure*}
	\includegraphics[scale=0.58,angle=0]{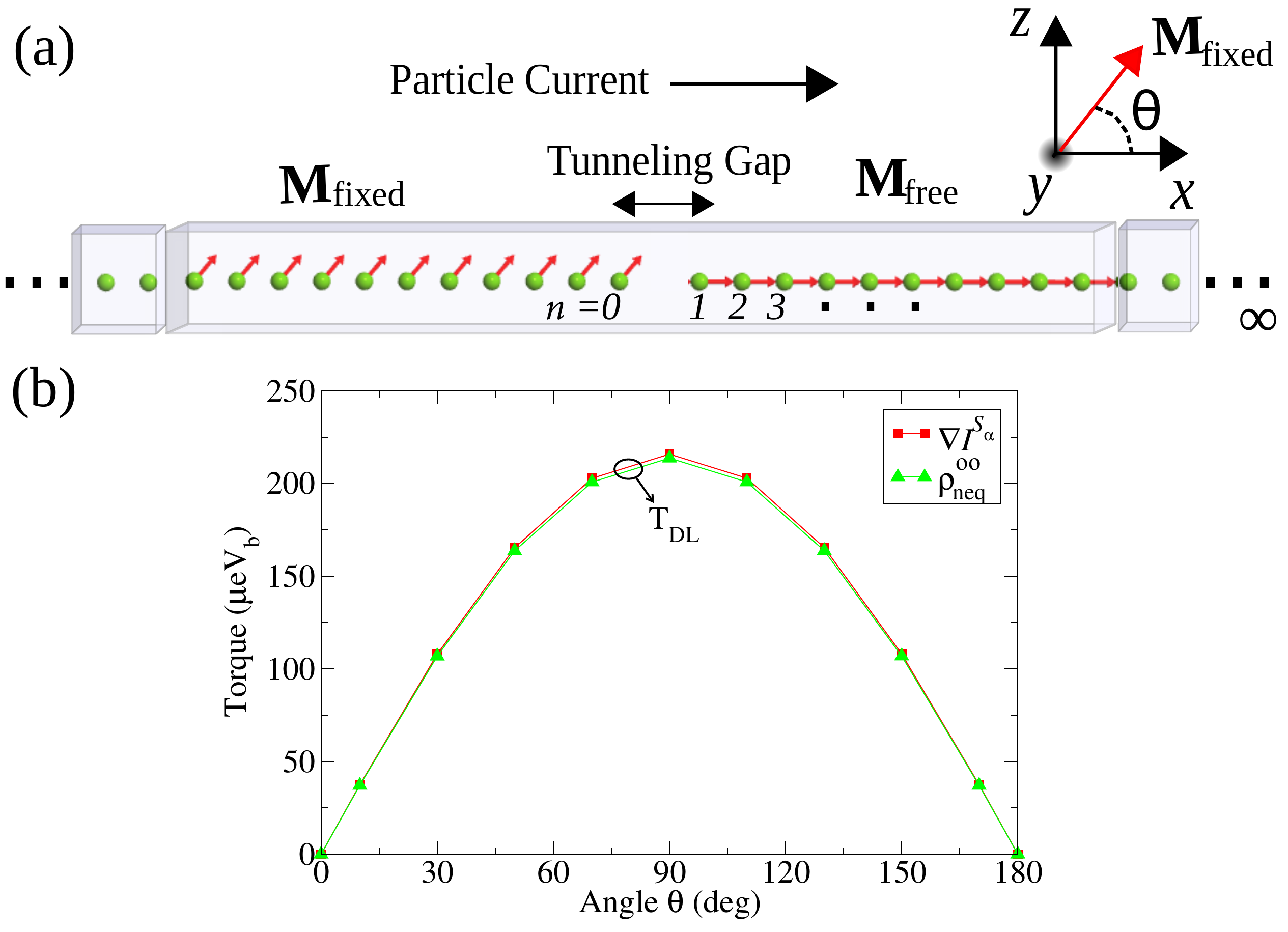}
	\caption{(a) Schematic view of a 1D toy model of MTJ consisting of the left and the right semi-infinite chains of carbon atoms separated by a vacuum gap. (b) Comparison of DL component of STT in this MTJ computed via the spin current divergence algorithm~\cite{Theodonis2006,Wang2008b} in Eq.~\eqref{eq:divergence} and by using decomposition of the nonequilibrium density matrix into ${\bm \rho}_\mathrm{neq}^{\mu \nu}$ contributions in Eqs.~\eqref{eq:rhooo}--\eqref{eq:rhoee}. The DL component of STT, computed in both algorithms in the linear-response regime, acts on the right carbon chain whose magnetic moments comprise the free magnetization $\mathbf{M}_\mathrm{free}$ rotated by an angle $\theta$ with respect to the fixed magnetization $\mathbf{M}_\mathrm{fixed}$ of the left carbon chain. Since MTJ is left-right symmetric, the FL component of the STT vector is zero in the linear-response regime~\cite{Wang2011,Theodonis2006,Xiao2008a,Oh2009,Heiliger2008}.}
	\label{fig:fig4}
\end{figure*}

In the calculations in Fig.~\ref{fig:fig3}, we first compute $(T^x_\mathrm{CD},T^y_\mathrm{CD},T^z_\mathrm{CD}) = \mathrm{Tr} \, [{\bm \rho}_{\rm CD} \mathbf{T}]$ using the torque operator $\mathbf{T}$ akin to Eq.~\eqref{eq:torqueop} but determined by the TBH of the free FM layer. These three numbers are then used to obtain FL (or perpendicular) torque component, $T_\mathrm{FL}=T^y_\mathrm{CD}$ along the direction $\mathbf{M}_\mathrm{free} \times  \mathbf{M}_\mathrm{fixed}$, and DL (or parallel) torque component, $T_\mathrm{DL}=\sqrt{(T^x_\mathrm{CD})^2 +(T^z_\mathrm{CD})^2}$ in the direction  $\mathbf{M}_\mathrm{free} \times (\mathbf{M}_\mathrm{free} \times \mathbf{M}_\mathrm{fixed})$. In MTJs angular dependence of STT components stems only from the cross product, so that  $\propto \sin \theta$ dependence~\cite{Theodonis2006,Xiao2008a} for both FL and DL components in obtained in Figs.~\ref{fig:fig3} and ~\ref{fig:fig4}.

In the case of SOT, $\mathbf{T}_\mathrm{DL} \propto \mathbf{M}_\mathrm{free} \times \mathbf{f}$ and $\mathbf{T}_\mathrm{DL} \propto \mathbf{M}_\mathrm{free} \times (\mathbf{M}_\mathrm{free} \times \mathbf{f})$, where the direction specified by the unit vector $\mathbf{f}$ is determined dynamically once the current flows in the presence of SOC. Therefore, $\mathbf{f}$ is not known in advance (aside from simplistic models like the Rashba ferromagnetic one where $\mathbf{f}$ is along the $y$-axis for charge current flowing along the $x$-axis, as illustrated in Fig.~\ref{fig:fig8}). Thus, it would be advantageous to decompose ${\bm \rho}_\mathrm{CD}$ into contributions whose trace with the torque operator in Eq.~\eqref{eq:torqueop} directly yield $\mathbf{T}_\mathrm{DL}$ and $\mathbf{T}_\mathrm{FL}$. Such decomposition was achieved in Ref.~\cite{Mahfouzi2016}, using adiabatic expansion of Eq.~\eqref{eq:rhoneqt} in the powers of $d \mathbf{M}_\mathrm{free}/dt$ and symmetry arguments, where ${\bm \rho}_\mathrm{neq} = {\bm \rho}_\mathrm{neq}^\mathrm{oo}+{\bm \rho}_\mathrm{neq}^\mathrm{oe}+{\bm \rho}_\mathrm{neq}^\mathrm{eo}+{\bm \rho}_\mathrm{neq}^\mathrm{ee}$ is the sum of the following terms
 \begin{eqnarray}
 {\bm \rho}^\mathrm{oo}_\mathrm{neq} & = & \frac{1}{8\pi} \int\limits_{-\infty}^{+\infty} dE \, \left[ f_L(E) - f_R(E) \right] \left(\mathbf{G} {\bm \Gamma}_L \mathbf{G}^\dagger - \mathbf{G}^\dagger {\bm \Gamma}_L \mathbf{G} -  \mathbf{G} {\bm \Gamma}_R \mathbf{G}^\dagger + \mathbf{G}^\dagger {\bm \Gamma}_R \mathbf{G} \right), \label{eq:rhooo} \\
 {\bm \rho}^\mathrm{oe}_\mathrm{neq} & = & \frac{1}{8\pi} \int\limits_{-\infty}^{+\infty} dE \, \left[ f_L(E) - f_R(E) \right] \left( \mathbf{G} {\bm \Gamma}_L \mathbf{G}^\dagger + \mathbf{G}^\dagger {\bm \Gamma}_L \mathbf{G} -  \mathbf{G} {\bm \Gamma}_R \mathbf{G}^\dagger - \mathbf{G}^\dagger {\bm \Gamma}_R \mathbf{G} \right), \label{eq:rhooe} \\
 {\bm \rho}^\mathrm{eo}_\mathrm{neq} & = & \frac{1}{8\pi} \int\limits_{-\infty}^{+\infty} dE \, \left[ f_L(E) + f_R(E) \right] \left(\mathbf{G} {\bm \Gamma}_L \mathbf{G}^\dagger - \mathbf{G}^\dagger {\bm \Gamma}_L \mathbf{G} +  \mathbf{G} {\bm \Gamma}_R \mathbf{G}^\dagger - \mathbf{G}^\dagger {\bm \Gamma}_R \mathbf{G} \right) \equiv 0, \label{eq:rhoeo} \\
 {\bm \rho}^\mathrm{ee}_\mathrm{neq} & = & \frac{1}{8\pi} \int\limits_{-\infty}^{+\infty} dE \, \left[ f_L(E) + f_R(E) \right] \left(\mathbf{G} {\bm \Gamma}_L \mathbf{G}^\dagger + \mathbf{G}^\dagger {\bm \Gamma}_L \mathbf{G} +  \mathbf{G} {\bm \Gamma}_R \mathbf{G}^\dagger + \mathbf{G}^\dagger {\bm \Gamma}_R \mathbf{G} \right). 
 \end{eqnarray}
The four terms are labeled by being odd (o) or even (e) under inverting bias polarity (first index) or time (second index)~\cite{Mahfouzi2016}. The terms  ${\bm \rho}^\mathrm{oo}_\mathrm{neq}$ and  ${\bm \rho}^\mathrm{oe}_\mathrm{neq}$ depend on $f_L(E) - f_R(E)$ and, therefore, are nonzero only in nonequilibrium generated by the bias voltage $V_b \neq 0$ which drives the steady-state current. Using an identity from the NEGF formalism~\cite{Stefanucci2013}, $\mathbf{G}({\bm \Gamma}_L + {\bm \Gamma_R})\mathbf{G}^\dagger = i(\mathbf{G}-\mathbf{G}^\dagger)$, reveals that ${\bm \rho}^\mathrm{eo}_\mathrm{neq} \equiv 0$ and 
\begin{equation}\label{eq:rhoee}
{\bm \rho}^\mathrm{ee}_\mathrm{neq} = -\frac{1}{2\pi} \int\limits_{-\infty}^{+\infty} dE\, [f_L(E) + f_R(E)] \mathrm{Im}\, \mathbf{G}.
\end{equation}
Thus, ${\bm \rho}^\mathrm{ee}_\mathrm{neq}$ term is nonzero even in equilibrium where it becomes identical to the equilibrium density matrix in Eq.~\eqref{eq:rhoeq},  $V_b = 0 \Rightarrow {\bm \rho}^\mathrm{ee}_\mathrm{neq} \equiv {\bm \rho}_\mathrm{eq}$. Since ${\bm \rho}^\mathrm{oo}_\mathrm{neq}$ is odd under time reversal, its trace with the torque operator in Eq.~\eqref{eq:torqueop} yields DL component of STT (which depends on three magnetization vectors and it is, therefore, also odd) and FL component of SOT (which depends on one magnetization vectors and it is, therefore, also odd). Similarly trace of ${\bm \rho}^\mathrm{oe}_\mathrm{neq}$ with the torque operator in Eq.~\eqref{eq:torqueop} yields FL component of STT and DL component of SOT~\cite{Mahfouzi2016}. 

In the linear-response regime, pertinent to calculations of STT in spin-valves and SOT in FM/spin-orbit-coupled-material bilayers, \mbox{$f_L(E)-f_R(E) \rightarrow (-\partial f/\partial E)eV_b$}. This confines  integration in ${\bm \rho}^\mathrm{oo}_\mathrm{neq}$ and ${\bm \rho}^\mathrm{oe}_\mathrm{neq}$ expressions to a shell of few $k_BT$ around the Fermi energy, or at zero temperature these are just matrix  products evaluated at the Fermi energy, akin to Eqs.~\eqref{eq:improper}, ~\eqref{eq:proper} and ~\eqref{eq:symmetricdm}. Nevertheless, to obtain ${\bm \rho}_\mathrm{CD}={\bm \rho}_\mathrm{neq}^\mathrm{oo}+{\bm \rho}_\mathrm{neq}^\mathrm{oe}+{\bm \rho}_\mathrm{neq}^\mathrm{eo}+{\bm \rho}_\mathrm{neq}^\mathrm{ee} - {\bm \rho}_\mathrm{eq}$ one still needs to perform the integration over the Fermi sea in order to obtain ${\bm \rho}_\mathrm{neq}^\mathrm{ee} - {\bm \rho}_\mathrm{eq}$, akin to Eq.~\eqref{eq:proper}, which can be equivalently computed as $[{\bm \rho}_\mathrm{neq}^\mathrm{ee}(V_b) - {\bm \rho}_\mathrm{neq}^\mathrm{ee}(-V_b)]/2$ using some small $V_b$. 

To evade singularities on the real axis caused by the poles of the retarded GF in the matrix integral of the type $\int\limits_{-\infty}^{+\infty} dE \, \mathbf{G} f_p(E)$ appearing in   Eqs.~\eqref{eq:rhoeq}, \eqref{eq:proper} and \eqref{eq:rhoee}, such integration can be performed along the contour in the upper half of the complex plane where the retarded GF is analytic. The widely used contour~\cite{Brandbyge2002} consists of a semicircle, a semi-infinite line segment and a finite number of poles of the Fermi function $f_p(E)$. This contour should be positioned sufficiently far away from the real axis, so that $\mathbf{G}$ is smooth over both of these two segments, while also requiring to select the minimum energy $E_\mathrm{min}$ (as the starting point of semicircular path)  below the bottom of the band edge which is not known in advance in DFT calculations. That is, in self-consistent calculations, incorrectly selected minimum energy  causes charge to erroneously disappear from the system with convergence trivially achieved but to physically incorrect solution. By choosing different types of contours~\cite{Areshkin2010,Ozaki2007,Karrasch2010} (such as the ``Ozaki contour''~\cite{Ozaki2007,Karrasch2010} we employ in the calculations in Fig.~\ref{fig:fig8}) where residue theorem leads to just a sum over a finite set of complex energies, proper positioning of $E_\mathrm{min}$ and convergence in the number of Fermi poles, as well as selection of sufficient number of contour points along the semicircle and contour points on the line segment are completely bypassed. 

Prior to application of the algorithm based on Eqs.~\eqref{eq:rhooo}--\eqref{eq:rhoee} to STT calculations in Sec.~\ref{sec:stt} and SOT calculations in Sec.~\ref{sec:sot}, we compare it to often employed spin current divergence algorithm~\cite{Theodonis2006,Wang2008b,Manchon2008a} using a toy model of 1D MTJ. The  model, illustrated in Fig.~\ref{fig:fig4}(a) where the left and right semi-infinite chains of carbon atoms are separated by a vacuum gap, is described by the collinear DFT Hamiltonian implemented in {\tt ATK} package~\cite{atk} using single-zeta polarized~\cite{Junquera2001} orbitals on each atom, Ceperly-Alder~\cite{Ceperley1980} parametrization of the local spin density approximation for the XC functional~\cite{Ceperley1980} and norm-conserving pseudopotentials accounting for electron-core interactions. In the absence of spin-flip processes by impurities and magnons or SOC, the STT vector at site $n$ within the right chain can be computed from the divergence (in discrete form) of spin current~\cite{Theodonis2006}, $T_n^\alpha = -\nabla I^{S_\alpha} =  I_{n-1,n}^{S_\alpha} -  I_{n,n+1}^{S_\alpha}$. Its sum over the whole free FM layer gives the total STT as
\begin{equation}\label{eq:divergence}
T^\alpha_\mathrm{CD} = \sum_{n=1}^{N_\mathrm{free}} (I_{n-1,n}^{S_\alpha} - I_{n,n+1}^{S_\alpha}) = I_{0,1}^{S_\alpha} - I_{N_\mathrm{free},N_\mathrm{free}+1}^{S_\alpha}.
\end{equation}
Here $I_{0,1}^{S_\alpha}$ is the local spin current, carrying spins pointing in the direction $\alpha \in \{x,y,z\}$, from the last site inside the barrier [which is the last site of the left carbon chain in Fig.~\ref{fig:fig4}(a)] toward the first site of the free FM layer [which is the first site of the right carbon chain in Fig.~\ref{fig:fig4}(a)]. Similarly,  $I_{N_\mathrm{free},N_\mathrm{free}+1}$ is the local spin current from the last site inside the free FM layer and  the first site of the right lead. Thus, Eq.~\eqref{eq:divergence} expresses STT on the  free FM layer composed of $N_\mathrm{free}$ sites as the difference~\cite{Wang2008b} between spin currents entering through its left and exiting through its right interface. In the case of semi-infinite free FM layer, $N_\mathrm{free} \rightarrow \infty$ and $I_{N_\mathrm{free},N_\mathrm{free}+1} \rightarrow 0$. The nonequilibrium local spin current can be computed in different ways~\cite{Wang2008b}, one of which utilizes NEGF expression for ${\bm \rho}_\mathrm{CD}$
\begin{equation}\label{eq:spincurrent}
I_{n,n+1}^{S_\alpha} =\frac{i}{2} \mathrm{Tr}\, \left[ {\bm \sigma}_\alpha \left(\mathbf{H}_{n,n+1} {\bm \rho}_\mathrm{CD}^{n+1,n}  - {\bm \rho}_\mathrm{CD}^{n,n+1} \mathbf{H}_{n+1,n} \right) \right].
\end{equation}
Here $\mathbf{H}_{n,n+1}$ and ${\bm \rho}_\mathrm{CD}^{n,n+1}$ are the submatrices of the Hamiltonian and the current-driven part of the nonequilibrium density matrix, respectively, of the size $2N_\mathrm{orbital} \times 2N_\mathrm{orbital}$ (2 is for spin and $N_\mathrm{orbital}$ is for the number of orbitals per each atom) which connect sites $n$ and $n+1$. 

Combining Eqs.~\eqref{eq:symmetricdm}, ~\eqref{eq:divergence} and ~\eqref{eq:spincurrent} yields $T^x_\mathrm{CD}$ and $T^z_\mathrm{CD}$ from which we obtain $T_\mathrm{DL}=\sqrt{(T^x_\mathrm{CD})^2 +(T^z_\mathrm{CD})^2}$ in the linear-response regime plotted in Fig.~\ref{fig:fig4}(b) as a function of angle $\theta$ between $\mathbf{M}_\mathrm{free}$ and $\mathbf{M}_\mathrm{fixed}$. Alternatively, evaluating the trace of the product of ${\bm \rho}_\mathrm{neq}^\mathrm{oo}$ and the torque operator in Eq.~\eqref{eq:torqueop} yields a vector with two nonzero components, which turn out to be identical to $T^x_\mathrm{CD}$ and $T^z_\mathrm{CD}$ computed from the spin current divergence algorithm, as demonstrated in Fig.~\ref{fig:fig4}(b). The trace of ${\bm \rho}_\mathrm{neq}^\mathrm{oe}$ with the torque operator gives a vector with zero $x$- and $z$-components and nonzero $y$-component which, however, is canceled by adding the trace of ${\bm \rho}_\mathrm{neq}^\mathrm{ee} - {\bm \rho}_\mathrm{eq}$ with the torque operator to finally produce zero FL component of the STT vector. This is expected because MTJ in Fig.~\ref{fig:fig4}(a) is left-right symmetric.

We emphasize that the algorithm based on the trace of the torque operator with the current-driven part of the nonequilibrium density matrix $\mathbf{\bm \rho}_\mathrm{CD}$ is a {\em more general approach} than the spin current divergence algorithm since it is valid even in the presence of spin-flip processes by impurities and magnons or SOC~\cite{Haney2007}. In particular, it can be employed to compute SOT~\cite{Freimuth2014,Mahfouzi2018} in FM/spin-orbit-coupled-material bilayers where spin torque {\em cannot}~\cite{Haney2010} be expressed any more as in Eq.~\eqref{eq:divergence}.

\section{Example: Spin-transfer torque in FM/NM/FM trilayer spin-valve structures}\label{sec:stt}

\begin{figure*}
	\includegraphics[scale=0.6,angle=0]{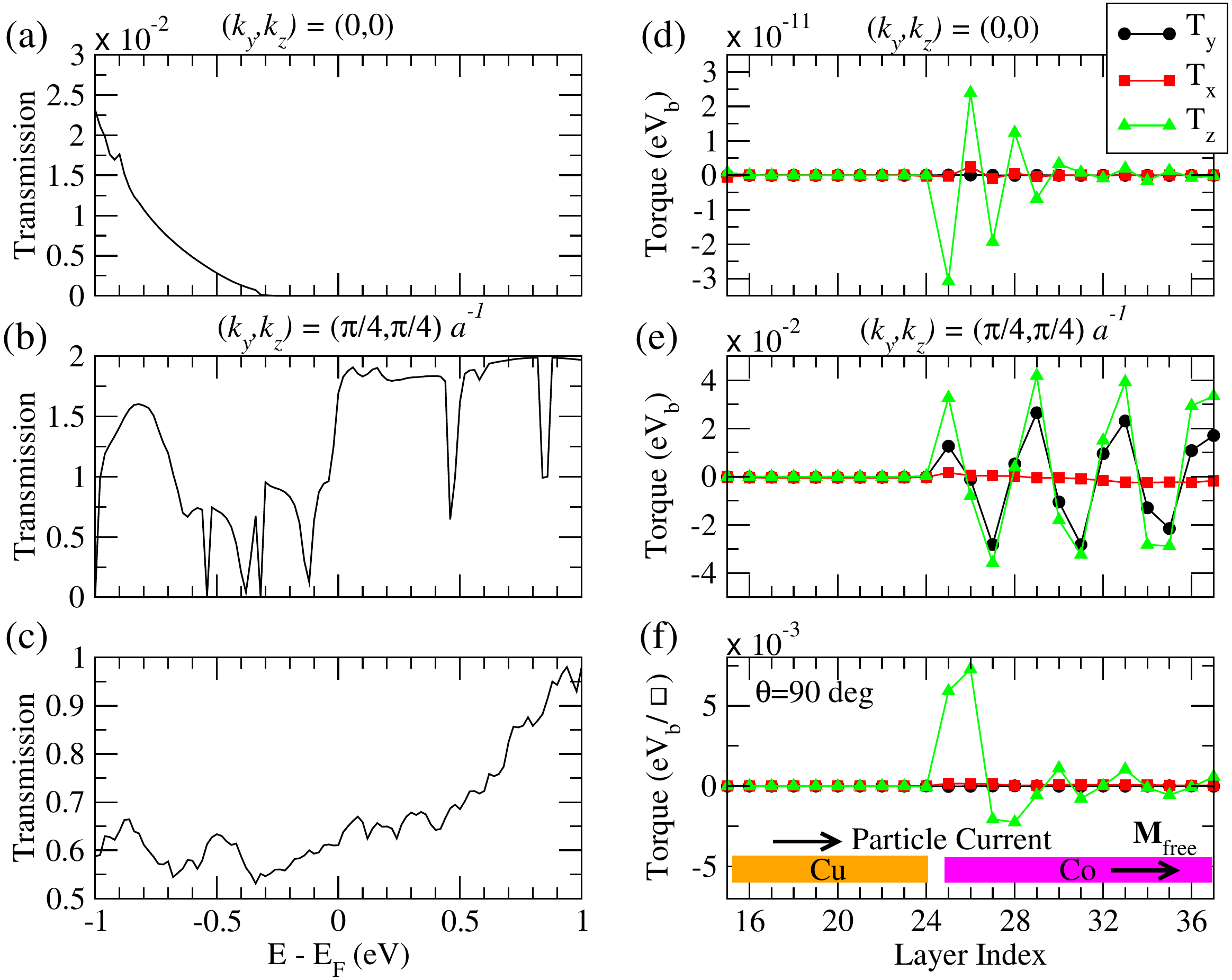}
	\caption{(a),(b) Transmission function, defined in Eq.~\eqref{eq:transmission}, for Co/Cu(9 ML)/Co spin-valve illustrated in Fig.~\ref{fig:fig1}(a) at two selected transverse wave vectors $\mathbf{k}_\parallel$; as well as (c) summed, as in Eq.~\eqref{eq:transmissionsum}, over all $\mathbf{k}_\parallel$ vectors within the 2D BZ. (d)--(f) The corresponding layer-resolved Cartesian components of the STT vector at an angle $\theta=90^\circ$ between the magnetizations $\mathbf{M}_\mathrm{fixed}$ of the left Co layer acting as a polarizer and $\mathbf{M}_\mathrm{free}$ of the right Co layer receiving the torque. The mesh of $\mathbf{k}_\parallel$ equally spaced points employed in (c) is $51 \times 51$, and in (f) it is $201 \times 201$. The symbol $\Box=a^2$ denotes unit interfacial area, where  $a =2.52$~\AA{} is the lattice constant of a common 2D hexagonal unit cell of Co/Cu bilayer (with a lattice mismatch of about 0.65\%).}
	\label{fig:fig5}
\end{figure*}

\begin{figure*}
	\includegraphics[scale=0.45,angle=0]{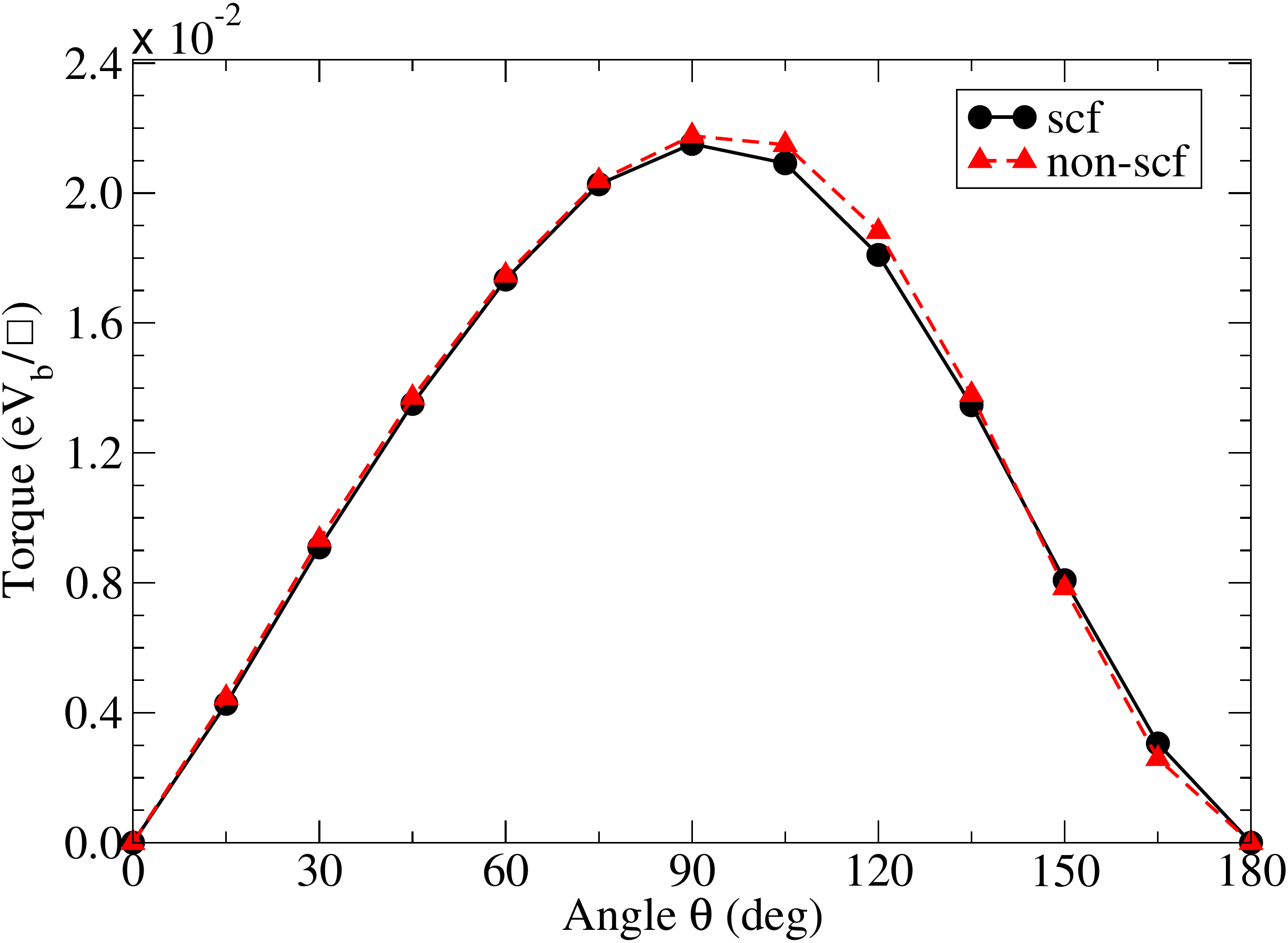}
	\caption{The DL component of STT vector in symmetric Co/Cu(9 ML)/Co spin-valve in Fig.~\ref{fig:fig1}(a) driven by current at small bias voltage $V_b$ as a function of angle $\theta$ between the magnetizations  $\mathbf{M}_\mathrm{fixed}$ of the left Co layer and $\mathbf{M}_\mathrm{free}$ of the right Co layer. Its value at $\theta=90^\circ$ is the sum of layer-resolved STT shown in Fig.~\ref{fig:fig5}(f). The non-scf curve is computed from Eq.~\eqref{eq:lcaotrace} using the rigid approximation~\cite{Haney2007,Wang2008b,Heiliger2008,Jia2011}, where spin-valve is described by collinear DFT Hamiltonian and  magnetic moments of the fixed Co layer are rotated by an angle theta with respect to magnetic moments in the free Co layer. Conversely, in the computation of scf curve, the DFT Hamiltonian from the rigid approximation serves as the first iteration leading toward converged self-consistent ncDFT Hamiltonian of the spin-valve.}
	\label{fig:fig6}
\end{figure*}

First-principles quantum transport modeling of STT in spin-valves~\cite{Haney2007,Wang2008b} and MTJs~\cite{Stamenova2017,Heiliger2008,Jia2011,Ellis2017} is typically conducted using an assumption that greatly simplifies computation---noncollinear spins in such systems are described in a {\em rigid} approximation where one starts from the collinear DFT Hamiltonian and then rotates magnetic moments of either fixed or free FM layer in the spin space in order to generate the relative angle between $\mathbf{M}_\mathrm{fixed}$ and $\mathbf{M}_\mathrm{free}$ (as it was also done in the calculations of STT in 1D toy model of MTJ in Fig.~\ref{fig:fig4}). On the other hand, obtaining true ground state of such system requires noncollinear XC functionals~\cite{Capelle2001,Eich2013a,Eich2013,Bulik2013} and the corresponding self-consistent XC magnetic field $\mathbf{B}_\mathrm{xc}$ introduced in Eq.~\eqref{eq:ncdft}. For a given self-consistently converged ncDFT Hamiltonian represented in the linear combination of atomic orbitals (LCAO) basis, we can extract the  matrix representation of $\mathbf{B}_\mathrm{xc}^\alpha$ in the same basis using
\begin{eqnarray}\label{eq:bxc}
\mathbf{B}_\mathrm{XC}^x &=& 2 \cdot {\rm Re} \, \mathbf{H}^{\uparrow \downarrow} , \\
\mathbf{B}_\mathrm{XC}^y &=& -2 \cdot {\rm Im} \, \mathbf{H}^{\uparrow \downarrow}, \\
\mathbf{B}_\mathrm{XC}^z &=&  \mathbf{H}^{\uparrow\uparrow} -  \mathbf{H}^{\downarrow\downarrow}. 
\end{eqnarray}
Since LCAO basis sets $|\phi_n \rangle$ are typically nonorthogonal (as is the case of the basis sets~\cite{Junquera2001,Ozaki2003,Schlipf2015} implemented in {\tt ATK}~\cite{atk} and {\tt OpenMX}~\cite{openmx} packages we employ in the calculations  of Figs.~\ref{fig:fig4}--\ref{fig:fig7}), the trace leading to the spin torque vector 
\begin{equation}\label{eq:lcaotrace}
\mathbf{T}_\mathrm{CD}=\mathrm{Tr}\, [{\bm \rho}_\mathrm{CD} {\bm \sigma} \times \mathbf{S}^{-1}\mathbf{B}_\mathrm{XC}],
\end{equation}
requires to use the identity operator $\mathbf{1} = \sum_{ij}|\phi_i\rangle S_{ij}^{-1} \langle \phi_j|$ which inserts $\mathbf{S}^{-1}$ matrix into Eq.~\eqref{eq:lcaotrace} where all matrices inside the trace are representations in the LCAO basis. In the real-space basis spanned by the eigenstates $|\mathbf{r} \rangle$  of the position operator, the same trace in Eq.~\eqref{eq:lcaotrace} becomes 
\begin{equation}\label{eq:realspacetrace}
\mathbf{T}_\mathrm{CD}=\int\limits_\mathrm{free \ FM} \!\!\! d^3\mathbf{r} \, \mathbf{m}_\mathrm{CD}(\mathbf{r}) \times \mathbf{B}_\mathrm{XC}(\mathbf{r}).
\end{equation}
This is a nonequilibrium generalization of the equilibrium torque expression found in ncDFT~\cite{Capelle2001}---effectively, $\mathbf{m}_\mathrm{eq}(\mathbf{r})=\langle \mathbf{r} |{\bm \rho}_\mathrm{eq} {\bm \sigma}|\mathbf{r} \rangle$ in ncDFT is replaced by \mbox{$\mathbf{m}_\mathrm{CD}(\mathbf{r}) = \langle \mathbf{r}| {\bm \rho}_\mathrm{CD} {\bm \sigma}| \mathbf{r} \rangle$}. Note that in thermodynamic equilibrium integral in Eq.~\eqref{eq:realspacetrace} must be zero when integration is performed over all space, which is denoted as ``zero-torque theorem''~\cite{Capelle2001}, but $\mathbf{m}_\mathrm{eq} \times \mathbf{B}_\mathrm{XC}(\mathbf{r})$ can be nonzero locally which gives rise to equilibrium torque on the free FM layer that has to be removed by subtracting ${\bm \rho}_\mathrm{eq}$ to obtain ${\bm \rho}_\mathrm{CD}$ in Eq.~\eqref{eq:rhocurr} and plug it into Eq.~\eqref{eq:realspacetrace}.

We employ {\tt ATK} package to compute STT in Co/Cu(9 ML)/Co spin-valve illustrated in Fig.~\ref{fig:fig1}(a) using ncDFT Hamiltonian combined with Eq.~\eqref{eq:lcaotrace}. Prior to DFT calculations, we employ the interface builder in the VNL package~\cite{vnl} to construct a common unit cell for Co/Cu bilayer. In order to determine the interlayer distance and relaxed atomic coordinates, we perform DFT calculations using {\tt VASP}~\cite{vasp,Kresse1993,Kresse1996a} with Perdew-Burke-Ernzerhof (PBE) parametrization~\cite{Perdew1996} of the generalized gradient approximation for the XC functional and projected augmented wave~\cite{Blochl1994,Kresse1999} description of electron-core interactions. The cutoff energy for the plane wave basis set is chosen as 600 eV, while $k$-points were sampled on a $11 \times 11$ surface mesh. In {\tt ATK} calculations we use PBE XC functional, norm-conserving pseudopotentials for describing electron-core interactions and SG15 (medium) LCAO basis set~\cite{Schlipf2015}. The energy mesh cut-off for the real-space grid is chosen as 100 Hartree.

The layer-resolved Cartesian components of STT vector within the free Co layer are shown in Fig.~\ref{fig:fig5}(d)--(f). The contribution from a propagating state oscillates as a function of position without decaying in Fig.~\ref{fig:fig5}(e) with a spatial period $2\pi/(k^\downarrow_\zeta - k^\uparrow_\eta)$ where  $\zeta$ ($\eta$) denotes different sheets~\cite{Wang2008b} of the Fermi surface for minority (majority) spin. This is due to the fact that noncollinear spin in Fig.~\ref{fig:fig1} entering right Co layer is not an eigenstate of the spin part of the  Hamiltonian determined by $\mathbf{M}_\mathrm{free}$ and it is, therefore, forced into precession. However, since the shapes of the Fermi surface for majority and minority spin in Co are quite different from each other~\cite{Wang2008b}, the spatial periods  of precession can vary rapidly for different $\mathbf{k}_\parallel$ within the 2D Brillouin zone (BZ). Thus, summation of their contributions leads to cancellation and, therefore, fast decay of STT away from the interface~\cite{Stiles2002,Wang2008b}, as demonstrated by plotting such sum to obtain the total 
STT per ML of Co in Fig.~\ref{fig:fig5}(f). 

The propagating vs. evanescent states are identified by finite [Fig.~\ref{fig:fig5}(b)] vs. vanishing [Fig.~\ref{fig:fig5}(a)], respectively, $\mathbf{k}_\parallel$-resolved transmission function obtained from the Landauer formula in terms of NEGFs~\cite{Stefanucci2013} 
\begin{equation}\label{eq:transmission}
T(E,\mathbf{k}_\parallel) = \mathrm{Tr}\, [{\bm \Gamma}_R(E,\mathbf{k}_\parallel) \mathbf{G}_0(E,\mathbf{k}_\parallel) {\bm \Gamma}_L(E,\mathbf{k}_\parallel) \mathbf{G}_0^\dagger(E,\mathbf{k}_\parallel)],
\end{equation}
where the transverse wave vector $\mathbf{k}_\parallel$ is conserved in the absence of disorder. The total transmission function per unit interfacial area is then evaluated using ($\Omega_\mathrm{BZ}$ is the area of sampled 2D BZ)
\begin{equation}\label{eq:transmissionsum}
T(E) = \frac{1}{\Omega_\mathrm{BZ}} \int\limits_\mathrm{2D \ BZ} T(E,\mathbf{k}_\parallel),
\end{equation}
as shown in Fig.~\ref{fig:fig5}(c). The transmission function in Fig.~\ref{fig:fig5}(a) at $\mathbf{k}_\parallel=(0,0)$ vanishes at the Fermi energy, signifying evanescent state which {\em cannot} carry any current across the junction. Nonetheless, such states can contribute~\cite{Ralph2008,Stiles2002,Wang2008b} to  STT vector, as shown in Fig.~\ref{fig:fig5}(d). Thus, the decay of STT away from Cu/Co interface in Fig.~\ref{fig:fig5}(f) arises both from the cancellation among contributions from propagating states with different $\mathbf{k}_\parallel$ and the decay of contributions from each evanescent state, where the latter are estimated~\cite{Wang2008b} to generate $\simeq 10$\% of the total torque on the ML of free Co layer that is closest to the Cu/Co interface in Fig.~\ref{fig:fig1}(a). 

Since the considered Co/Cu(9 ML)/Co spin-valve is left-right symmetric, the FL component of the STT vector is zero. The DL component, as the sum of all layer-resolved torques in Fig.~\ref{fig:fig5}(f), is plotted as a function of the relative angle $\theta$ between $\mathbf{M}_\mathrm{fixed}$ and $\mathbf{M}_\mathrm{free}$ in Fig.~\ref{fig:fig6}. The angular dependence of STT in spin-valves does not follow $\propto \sin \theta$ dependence found in the case of MTJs in Figs.~\ref{fig:fig3} and ~\ref{fig:fig4}.

Although similar analyses have been performed before using collinear DFT Hamiltonian and rigid rotation of magnetic moments in fixed Co layer~\cite{Haney2007,Wang2008b}, in Fig.~\ref{fig:fig6} we additionally quantify an error made in this approximation by computing torque using ncDFT Hamiltonian. The rigid approximation is then just the first iteration of the full self-consistent field (scf) calculations leading to our converged ncDFT Hamiltonian. The difference between scf and non-scf calculations in Fig.~\ref{fig:fig6} is rather small due to large number of spacer MLs of Cu, but it could become sizable for small number of spacer MLs enabling coupling between two FM layers.

\section{Example: Spin-orbit torque in FM/monolayer-TMD bilayers}\label{sec:sot}

The calculation of SOT driven by injection of unpolarized charge current into bilayers of the type FM/HM shown in Fig.~\ref{fig:fig1}(b); FM/monolayer-TMD shown in Fig.~\ref{fig:fig1}(c);  FM/TI; or FM/WSM can be performed using the same NEGF+ncDFT framework combining the torque operator $\mathbf{T}$, ${\bm \rho}_\mathrm{CD}$ expressed in terms of NEGFs and ncDFT Hamiltonian that was delineated in Sec.~\ref{sec:dm} and applied in Sec.~\ref{sec:stt} to compute the STT vector in spin-valves. Such first-principles quantum transport approach can also easily accommodate  possible third capping insulating layer (such as MgO or AlO$_x$) employed experimentally to increase~\cite{Kim2013} the perpendicular magnetic anisotropy which tilts magnetization out of the plane of the interface. However, the results of such calculations are not as easy to interpret as in the case of a transparent picture~\cite{Stiles2002,Wang2008b} in  Fig.~\ref{fig:fig5}(d)--(f) explaining how spin angular momentum gets absorbed close to the interface in junctions which exhibit conventional STT. This is due to the fact that several microscopic mechanisms can contribute to SOT, such as: the spin Hall effect (SHE)~\cite{Vignale2010,Sinova2015} within the HM layer~\cite{Freimuth2014,Mahfouzi2018} with strong bulk SOC and around FM/HM interface~\cite{Wang2016a}; current-driven nonequilibrium spin density---the so-called Edelstein effect~\cite{Edelstein1990,Aronov1989}---due to strong interfacial SOC; spin currents generated in transmission and reflection from SO-coupled interfaces within 3D transport  geometry~\cite{Zhang2015,Kim2017}; and spin-dependent scattering off impurities~\cite{Pesin2012a,Ado2017} or  boundaries~\cite{Mahfouzi2016} in the presence of SOC within FM monolayers. This makes it difficult to understand how to optimize SOT by tailoring materials combination or device geometry to enhance one or more of these mechanisms. 

\begin{figure*}
	\includegraphics[scale=0.45,angle=0]{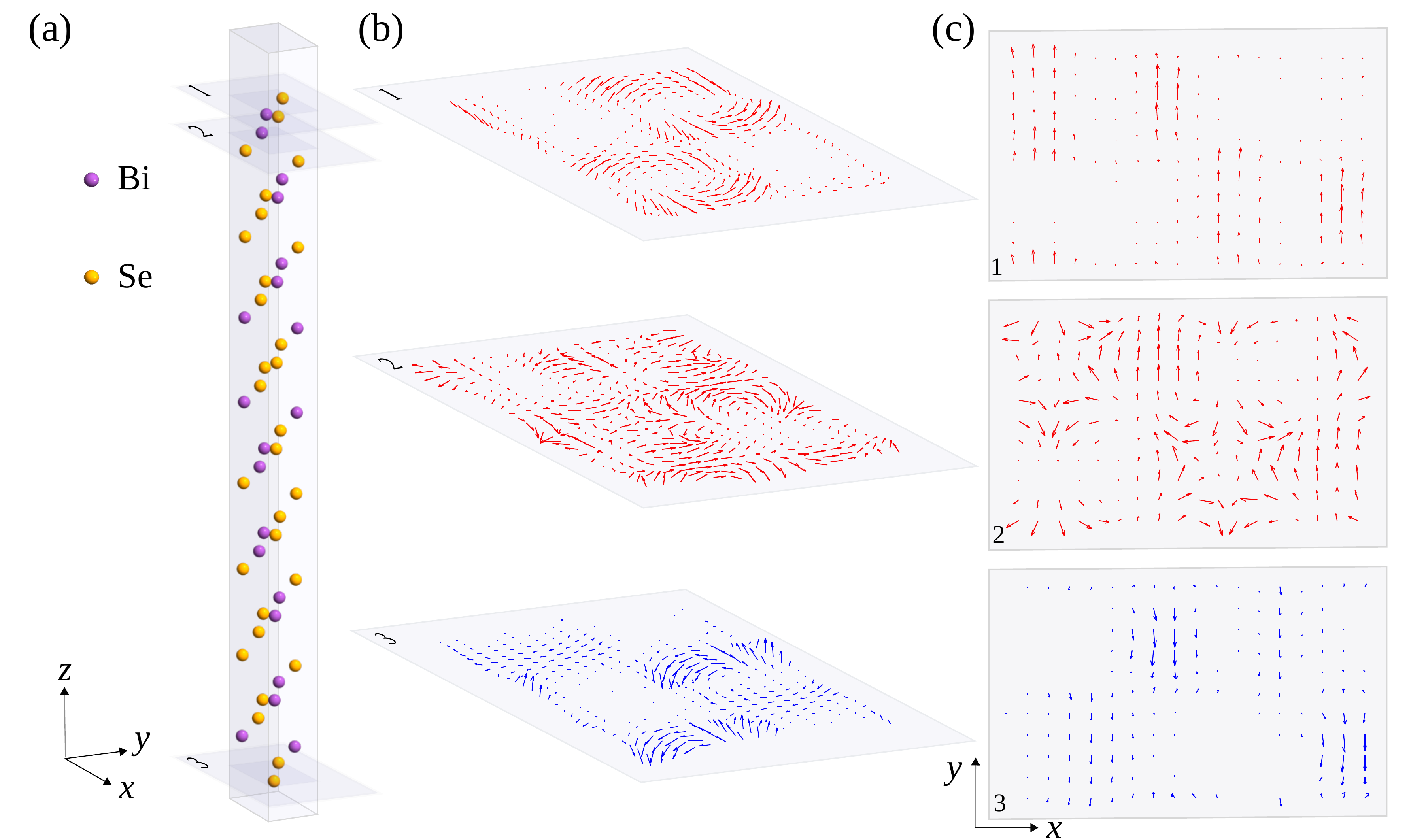}
	\caption{(a) The arrangement of Bi and Se atoms within a supercell of Bi$_2$Se$_3$ thin film (infinite in the $x$- and the $y$-directions) consisting of five quintuple layers (one such  layer contains three Se layers strongly bonded to two Bi layers in between) and total thickness \mbox{$\approx 5$ nm}.  (b) The vector field of current-driven nonequilibrium spin density $\mathbf{S}_\mathrm{CD}(\mathbf{r})$ within selected planes shown in (a), generated by injection of unpolarized charge current along the $x$-axis. The planes 1 and 3 correspond to the top and bottom metallic surfaces of Bi$_2$Se$_3$ thin film, while plane 2 resides in the bulk at a distance \mbox{$d \approx 0.164$ nm} away from plane 1. (c) The vector fields in (b) projected onto each of the selected planes in (a). The real space grid of points in panels (b) and (c) has spacing $\simeq 0.4$ \AA{}. Adapted from Ref.~\cite{Chang2015}.}
	\label{fig:fig7}
\end{figure*}

An example of first-principles quantum transport modeling of the Edelstein effect is shown in Fig.~\ref{fig:fig7} for the case of the metallic surface of Bi$_2$Se$_3$ as the prototypical 3D TI~\cite{Bansil2016}. Such materials possess a usual band gap in the bulk, akin to conventional topologically trivial insulators, but they also host metallic surfaces whose low-energy quasiparticles behave as massless Dirac fermions. The spins of such fermions are perfectly locked to their momenta by strong SOC, thereby forming spin textures in the reciprocal-space~\cite{Bansil2016}. In general, when charge current flows through a surface or interface with SOC, the presence of SOC-generated spin texture in the reciprocal-space, such as those shown in Figs.~\ref{fig:fig2}(e),  ~\ref{fig:fig2}(g) and ~\ref{fig:fig2}(h), will generate nonequilibrium spin density which can be computed using 
\begin{equation}\label{eq:sneq}
\mathbf{S}_\mathrm{CD} = \frac{\hbar}{2} \mathrm{Tr}\, [{\bm \rho}_\mathrm{CD} {\bm \sigma}].
\end{equation}
In the case of simplistic Hamiltonians---such as the Rashba one for 2D electron gas~\cite{Winkler2003}, $\hat{H}_\mathrm{Rashba}=(\hat{p}_x^2+\hat{p}_y^2)/2m^* + \frac{\alpha_\mathrm{SO}}{\hbar}(\hat{\sigma}_x \hat{p_y} - \hat{\sigma}_y \hat{p_x})$, or the Dirac one for the metallic surface of 3D TI, $\hat{H}_\mathrm{Dirac}= v_F (\hat{\sigma}_x \hat{p_y} - \hat{\sigma}_y \hat{p_x})$---the direction and the magnitude of $\mathbf{S}_\mathrm{CD}$ are easily determined by back-of-the-envelope calculations~\cite{Pesin2012}. For example, spin texture (i.e., expectation value of the spin operator in the eigenstates of a Hamiltonian) associated with $\hat{H}_\mathrm{Rashba}$ consists of spin vectors locked to momentum vector along the two Fermi circles formed in the reciprocal-space at the intersection of the Rashba energy-momentum dispersion~\cite{Winkler2003} and the Fermi energy plane. Thus, current flow will disturb balance of momenta to produce $\mathbf{S}_\mathrm{CD}$ in the direction transverse to current flow. The same effect is substantially enhanced~\cite{Pesin2012}, by a factor $\hbar v_F/\alpha_\mathrm{SO} \gg 1$ where $v_F$ is the Fermi velocity in TI and $\alpha_\mathrm{SO}$ is the strength of the Rashba SOC, because spin texture associated with $\hat{H}_\mathrm{Dirac}$ consists of spin vectors locked to momentum vector along a {\em single} Fermi circle formed in the reciprocal-space at the intersection of the Dirac cone energy-momentum dispersion~\cite{Bansil2016} and the Fermi energy plane. This eliminates the compensating effect of the spins along the second circle in the case of $\hat{H}_\mathrm{Rashba}$. Note that nonzero total $\mathbf{S}_\mathrm{CD} \propto V_b$ generated by  the Edelstein effect is allowed only in nonequilibrium since in equilibrium $\mathbf{S}$ changes sign under time reversal, and, therefore, has to vanish (assuming absence of external magnetic field or magnetization). 

In the case of a thin film of Bi$_2$Se$_3$ described by ncDFT Hamiltonian, unpolarized injected charge current generates $\mathbf{S}_\mathrm{CD} = (0,S^y_\mathrm{CD},S^z_\mathrm{CD})$ on the top surface of the TI, marked as plane 1 in Fig.~\ref{fig:fig7}, where the in-plane component $S^y_\mathrm{CD}$  is an order of magnitude larger than the ouf-of-plane component $S^z_\mathrm{CD}$ arising due to hexagonal warping~\cite{Bansil2016} of the Dirac cone on the TI surface. The spin texture on the bottom surface of the TI, marked as plane 3 in Fig.~\ref{fig:fig7}, has opposite sign to that shown on the top surface because of opposite direction of spins wounding along single Fermi circle on the bottom surface. In addition, a more complicated spin texture in real space (on a grid of points with $\simeq 0.4$ \AA{} spacing), akin to noncollinear intra-atomic magnetism~\cite{Nordstroem1996} but driven here by current flow, emerges within $\simeq 2$ nm thick layer below the TI surfaces. This is due to the penetration of evanescent wave functions from the metallic surfaces into the bulk of the TI, as shown in Fig.~\ref{fig:fig7} by plotting $\mathbf{S}_\mathrm{CD}(\mathbf{r})$ within plane 2. 

The conventional unpolarized charge current injected into the HM layer in Fig.~\ref{fig:fig1}(b) along the $x$-axis generates transverse spin Hall currents~\cite{Vignale2010,Sinova2015,Wang2016a} due to strong SOC in such layers. In 3D geometry, spin Hall current along the $y$-axis carries spins polarized along the $z$-axis, while the spin Hall current along the $z$-axis carries spins polarized along the $y$-axis~\cite{Wang2016a}. Thus, the effect of the spin Hall current flowing along the $z$-axis and entering FM layer resembles STT that would be generated by a fictitious polarizing FM layer with fixed magnetization along the $y$-axis and with charge current injected along the $z$-axis. While this mechanism is considered to play a major role in the generation of the DL component of SOT~\cite{Liu2012c}, as apparently confirmed by the Kubo-formula+ncDFT modeling~\cite{Freimuth2014,Mahfouzi2018}, the product of signs of the FL and DL torque components is negative in virtually all experiments~\cite{Yoon2017} (except for specific thicknesses of HM=Ta~\cite{Kim2013} layer). Although HM layer in  Fig.~\ref{fig:fig1}(b) certainly generates spin Hall current in its bulk, such spin current can be largely suppressed by the spin memory loss~\cite{Dolui2017,Belashchenko2016} as electron traverses HM/FM  interface with strong SOC. Thus, in contrast to positive sign product in widely accepted picture where  SHE is most responsible for the DL component of SOT and Edelstein effect is most responsible for the FL component of SOT, negative sign product indicates that a single dominant mechanism could be responsible for {\em both} DL and FL torque. 

\begin{figure*}
	\includegraphics[scale=1.0,angle=0]{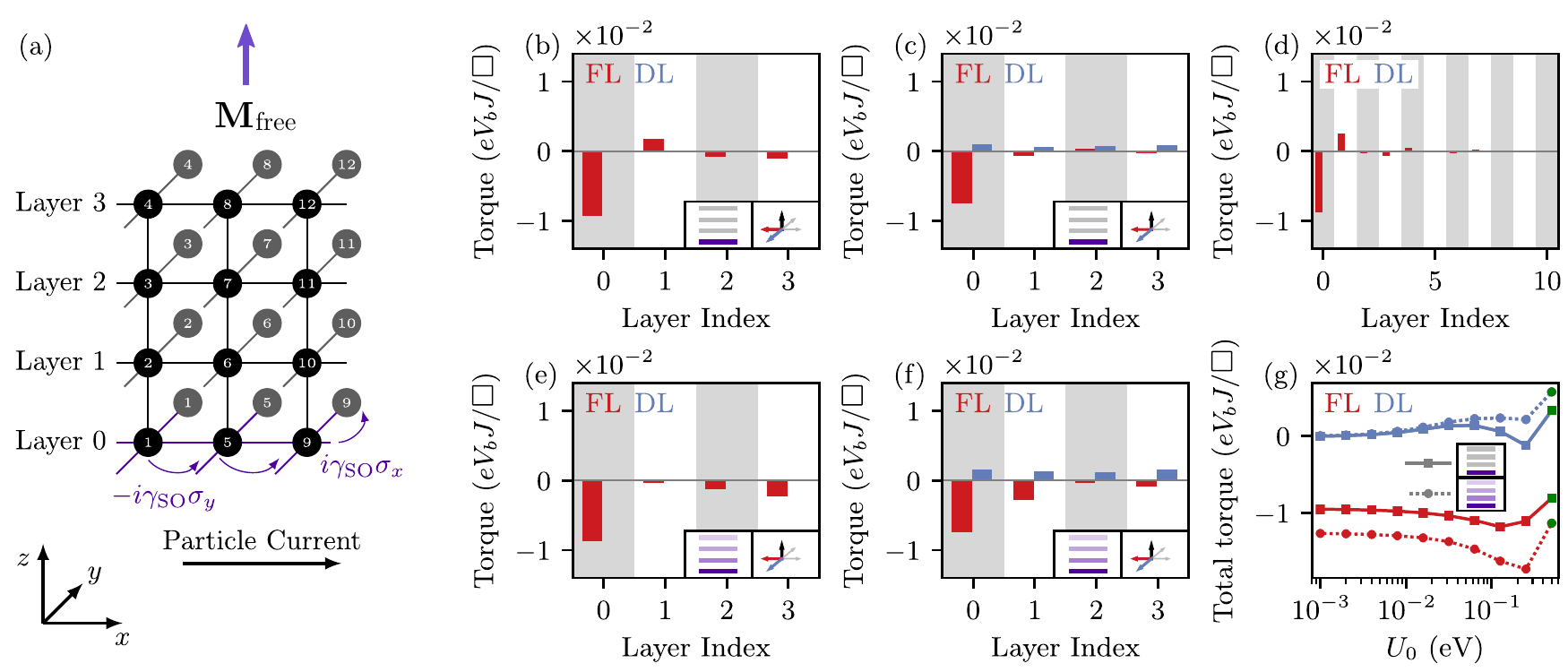}
	\caption{(a) Schematic view of an ultrathin FM layer comprised of few MLs, modeled as infinite square tight-binding lattices (with lattice constant $a$) in the $xy$-plane. The bottom layer 0 is assumed to be in direct contact with a spin-orbit-coupled material like TMD.  (b)---(g) The layer-resolved FL and DL components of SOT vector per unit area $\Box=a^2$ for the device in (a) where FM film is 4 MLs thick in all panels except (d) where the thickness is 10 MLs.  The bottom layer 0 is assumed to host the Rashba SOC of strength $\gamma_\mathrm{SO}=0.2$ eV, induced by the spin-orbit proximity effect  from monolayer-TMD (which is assumed to be insulating and, therefore, not explicitly considered), as illustrated in Fig.~\ref{fig:fig2}(h) for Co/monolayer-MoS$_2$. This is twice as strong as the exchange field in the FM layer, $\gamma_\mathrm{SO}/J=2$ in Eq.~\eqref{eq:fmhamiltonian}. In panels (e) and (f) we also add the Rashba SOC in layers $l>0$ with strength 50\% smaller than in layer $l-1$ (i.e., $\gamma_\mathrm{SO}=0.2$ eV, $0.1$ eV, $0.05$ eV and $0.025$ eV in layers 0--3, respectively). Panel (g) plots FL and DL components of SOT summed over all layers as a function of the strength of a homogeneous on-site potential $U_\mathbf{n}=U_0$ in Eq.~\eqref{eq:onsite} added into the central region of the thickness of 5 sites along the $x$-axis. The green square and dot in panel (g) mark SOT components at $U_0=0.5$ eV whose layer resolved values are shown in panels (c) and (f), respectively. The temperature is set at $T = 300$~K, Fermi energy is  $E_F = 1$~eV and we use a grid of $N_{k_y}=2000$ equally spaced $k_y$-points to sample periodically repeated lattice in the $y$-direction. The red and blue arrows in the insets of panels (b),(c),(e) and (f) denote $\mathbf{M}_\mathrm{free} \times \mathbf{e}_y$ and $\mathbf{M}_\mathrm{free} \times (\mathbf{M}_\mathrm{free} \times \mathbf{e}_y)$ unit vectors along the direction of FL and DL components of SOT, respectively. The quad-graphs in the insets of panels (b), (c), (e), (f) and (g) signify presence or absence of the Rashba SOC within layers 0--3 using the shade of the corresponding line.}
	\label{fig:fig8}
\end{figure*}

In order to explore how such single mechanism could  arise in the absence of spin Hall current, one can calculate torque in a number of specially crafted setups, such as the one chosen in Fig.~\ref{fig:fig8}(a) where we consider a single ultrathin FM layer  (consisting of 4 or 10 MLs) with the Rashba SOC present either only in the bottom ML [marked as 0 in Fig.~\ref{fig:fig8}(a)] or in all MLs but with decreasing strength to mimic the spin-orbit  proximity effect exemplified in Figs.~\ref{fig:fig2}(g) and ~\ref{fig:fig2}(h). The setup in  Fig.~\ref{fig:fig8}(a) is motivated by SOT experiments~\cite{Sklenar2016,Shao2016,Guimaraes2018,Lv2018} on FM/monolayer-TMD heterostructures where SHE in absent due to atomically thin spin-orbit-coupled material. In such bilayers, clean and atomically precise interfaces have been achieved while back-gate voltage~\cite{Lv2018} has been employed to demonstrate control of the ratio between the FL and DL components of SOT. Note that when bulk TMD or its even-layer thin films are centrosymmetric, its monolayer will be noncentrosymmetric crystal which results in lifting of the spin degeneracy and possibly strong SOC effects~\cite{Zhu2011a}.

In order to be able to controllably switch SOC on and off in different MLs, or to add other types of single-particle interactions, we describe setup in Fig.~\ref{fig:fig8}(a) using TBH defined on an infinite cubic lattice of spacing $a$ with a single orbital per site located at position $\mathbf{n} = (n_{x}a, n_{z}a)$
\begin{equation}\label{eq:fmhamiltonian}
\hat{H}  =  \sum_{\mathbf{n}} \varepsilon_{\mathbf{n}} \hat{c}^\dagger_{\mathbf{n}} \hat{c}_{\mathbf{n}} +\sum_{\langle \mathbf{n} \mathbf{n}'\rangle}  \left(\hat{c}^\dagger_{\mathbf{n}} \mathbf{t}_{\mathbf{n}\mathbf{n}'} \hat{c}_{\mathbf{n'}} + {\rm h.c.}\right) -  
J \sum_{\mathbf{n}}  \hat{c}^\dagger_{\mathbf{n}} \mathbf{M}_\mathrm{free} \cdot{\boldsymbol \sigma} \hat{c}_{\mathbf{n}}.
\end{equation}
Here $\hat{c}_\mathbf{n}^\dagger=(\hat{c}^\dagger_{\mathbf{n}\uparrow} \ \ \hat{c}^\dagger_{\mathbf{n}\downarrow})$ is the row vector of operators which create electron at site $\mathbf{n}$ in spin-$\uparrow$ or spin-$\downarrow$ state, and $\hat{c}_\mathbf{n}=(\hat{c}_{\mathbf{n}\uparrow} \ \ \hat{c}_{\mathbf{n}\downarrow})$ is the column vector of the corresponding annihilation operators. The spin-dependent nearest-neighbor hoppings in the $xz$-plane form a $2 \times 2$ matrix in the spin space
\begin{equation}\label{eq:hopping}
\mathbf{t}_{\mathbf{n}\mathbf{n}'} = \left\{
\begin{array}{ll}
-t\otimes I_{2}, &
{\rm for }\ \ \mathbf{n} = \mathbf{n}' +
{\boldsymbol e}_{z},\\
-t\otimes I_{2} - i\gamma_{\rm SO}\sigma_y, &
{\rm for }\ \ \mathbf{n} = \mathbf{n}' +
{\boldsymbol e}_{x},\\
\end{array}
\right.
\end{equation}
where $\gamma_\mathrm{SO}=\alpha_\mathrm{SO}/2a$ measures the strength of the Rashba SOC on the lattice~\cite{Nikolic2006}, $I_2$ is unit $2 \times 2$ matrix and $\mathbf{e}_\alpha$ are the unit vectors along the axes of the Cartesian coordinate system. The on-site energy  
\begin{equation}\label{eq:onsite}
\varepsilon_{\mathbf{n}} = \left[U_\mathbf{n} - 2t\cos (k_y a\right)] \otimes I_{2} - 2\gamma_{\rm SO } \sigma_x \sin (k_y a),
\end{equation}
includes the on-site potential energy $U_\mathbf{n}$ (due to impurities, insulating barrier, voltage drop, etc.), as well as kinetic energy effectively generated by the periodic 
boundary conditions along the $y$-axis which simulate infinite extension of the FM layer in this direction and require $k_y$-point sampling in all calculations. The infinite extension along the $x$-axis is taken into account by splitting the device in Fig.~\ref{fig:fig8}(a) into semi-infinite left lead, central region of arbitrary length along the $x$-axis, and semi-infinite right lead, all of which are described by the Hamiltonian in Eq.~\eqref{eq:fmhamiltonian} with the same values for $t=1$ eV, $J=0.1$ eV and the $\gamma_\mathrm{SO}$ chosen in all three regions. Thus,  $\gamma_\mathrm{SO}$ is homogeneous within a given ML, and always present $\gamma_\mathrm{SO}=0.2$ eV on layer 0, but it can take different values in other MLs. The Fermi energy is set at $E_F=1.0$ eV to take into account possible noncircular Fermi surface~\cite{Lee2015} effects in realistic materials.

The SOT is often studied~\cite{Manchon2008,Haney2013,Lee2015,Li2015,Pesin2012a,Ado2017,Kalitsov2017} using the Rashba ferromagnetic model in 2D, which corresponds to just a single layer in Fig.~\ref{fig:fig8}(a). In that case, only FL torque component is found~\cite{Manchon2008,Kalitsov2017} due to the Edelstein effect, and in the absence of spin-dependent disorder~\cite{Pesin2012a,Ado2017}. This is also confirmed is our 3D transport geometry in Figs.~\ref{fig:fig8}(b), ~\ref{fig:fig8}(d) and ~\ref{fig:fig8}(e) where SOT vector, $\frac{2J}{\hbar} \mathbf{S}_\mathrm{CD} \times \mathbf{M}_\mathrm{free}$, on layer 0 has only FL component pointing in the $\mathbf{M}_\mathrm{free} \times \mathbf{e}_y$ direction, as long as the device is infinite clean and homogeneous. In addition, we also find nonzero FL component of SOT in Figs.~\ref{fig:fig8}(b) and ~\ref{fig:fig8}(d) on layers above layer 0 despite that fact that only layer 0 hosts $\gamma_\mathrm{SO} \neq 0$. This is due to vertical transport along the $z$-axis in 3D geometry of Fig.~\ref{fig:fig8}(a), but in the absence of the Rashba SOC on other layers such effect  decays fast as we move toward the top layer, as shown in Fig.~\ref{fig:fig8}(d) for 10 MLs thick FM film. The presence of the Rashba SOC with decreasing $\gamma_\mathrm{SO}$ on MLs above layer 0 generates additional nonequilibrium spin density $\mathbf{S}_\mathrm{CD} \propto \mathbf{e}_y$ on those layers and the corresponding enhancement of the FL component of SOT  on those layers in ~\ref{fig:fig8}(e). 

We note that the Kubo-Bastin formula~\cite{Bastin1971} adapted for SOT calculations~\cite{Freimuth2014} predicts actually nonzero DL component of SOT vector for the Rashba ferromagnetic model in 2D due to change of electronic wave functions induced by an applied electric field termed the ``Berry curvature mechanism''~\cite{Lee2015,Kurebayashi2014,Li2015}. Despite being apparently intrinsic, i.e., insensitive to disorder, this mechanism can be completely canceled in specific models when the vertex corrections are taken into account~\cite{Ado2017}. It also gives positive sign product~\cite{Lee2015,Li2015} of DL and FL components of SOT contrary to majority of experiments where such product is found to be negative~\cite{Yoon2017}. We emphasize that no electric field can exist in the ballistic transport regime through clean devices we analyze in Figs.~\ref{fig:fig8}(b), ~\ref{fig:fig8}(d) and ~\ref{fig:fig8}(e), for which the Kubo-Bastin formula also predicts unphysical divergence~\cite{Freimuth2014,Mahfouzi2018,Lee2015,Kurebayashi2014,Li2015} of the FL component of SOT. Adding finite voltage drop within the central region, which is actually unjustified in the case of infinite clean homogeneous device, results in nonzero DL component of SOT also in the NEGF calculations~\cite{Kalitsov2017}. However, the same outcome can be obtained simply by introducing constant potential $U_\mathbf{n} = U_0$ on each site of the central region acting as a barrier which reflects incoming electrons, as demonstrated in Figs.~\ref{fig:fig8}(c), ~\ref{fig:fig8}(f) and ~\ref{fig:fig8}(g). In the presence of both SOC and such barrier, spin-dependent scattering~\cite{Pesin2012a,Ado2017}  is generated at the lead/central-region boundary which results in nonzero component of $\mathbf{S}_\mathrm{CD}$ in the direction $\mathbf{M}_\mathrm{free} \times \mathbf{e}_y$ and the corresponding DL component of SOT $\propto \mathbf{M}_\mathrm{free} \times (\mathbf{M}_\mathrm{free} \times \mathbf{e}_y)$ acting on edge magnetic moments~\cite{Mahfouzi2016}. This will, therefore, induce inhomogeneous magnetization switching which starts from the edges and propagates into the bulk of FM layer, as observed in experiments~\cite{Baumgartner2017} and micromagnetic simulations~\cite{Baumgartner2017,Mikuszeit2015}. 

Interestingly, Fig.~\ref{fig:fig8}(g) also demonstrates that the signs of the DL and FL component are opposite to each other for almost all values of $U_\mathbf{n} =U_0$ (except when $U_0 \simeq J$), as observed in majority of SOT experiments~\cite{Yoon2017}. Importantly, the spin-orbit proximity effect within the MLs of FM layer close to FM/spin-orbit-coupled-material interface, as  illustrated in Figs.~\ref{fig:fig2}(g) and ~\ref{fig:fig2}(h) and mimicked by introducing the Rashba SOC of decaying strength within all MLs of FM thin film in Fig.~\ref{fig:fig8}(a), enhances both FL and DL components of SOT. This is demonstrated by comparing solid ($\gamma_\mathrm{SO} \neq 0$ only on layer 0) and dashed ($\gamma_\mathrm{SO} \neq 0$ on all layers 0--3) lines in Fig.~\ref{fig:fig8}(g). This points out at a knob that can be exploited to enhance SOT by searching for optimal combination of materials capable to generate penetration of SOC over long distances within the FM layer~\cite{Marmolejo-Tejada2017}. In fact, in the case of FM/TI and FM/monolayer-TMD heterostructures, proximity SOC coupling within the FM layer is crucial for SOT efficiency~\cite{Wang2017}  where it has been considered~\cite{Mellnik2014} that applied current will be shunted through the metallic FM layer and, therefore, not contribute to nonequilibrium spin density generation at the interface where SOC and thereby induced in-plane spin textures are naively assumed to reside.

\section{Conclusions}\label{sec:conclusions}
This Chapter reviews a unified first-principles quantum transport approach, implemented by combining NEGF formalism with ncDFT calculations, to compute both STT in traditional magnetic multilayers with two FM layers (i.e., the polarizing and analyzing FM layers with fixed and free magnetizations, respectively) and SOT in magnetic bilayers where only one of the layers is ferromagnetic. In the latter case, the role of the fixed magnetization of the polarizing FM layer within spin-valves or MTJs is taken over by the current-driven nonequilibrium spin density in the presence of strong SOC introduced by the second layer made of HM, 3D TI, WSM or monolayer-TMD. Our approach resolves recent confusion~\cite{Freimuth2014,Kalitsov2017} in the literature where apparently only the Kubo formula, operating with expressions that integrate over the Fermi sea in order to capture change of wave functions due to the applied electric field and the corresponding interband electronic transitions~\cite{Freimuth2014,Lee2015,Li2015}, can properly obtain the DL component of SOT. In addition, although the Kubo formula approach can also be integrated with first-principles calculations~\cite{Freimuth2014,Mahfouzi2018}, it can only be applied to a single device geometry (where infinite FM layer covers infinite spin-orbit-coupled-material layer while current flows parallel to their interface) and in the linear-response transport regime. In contrast, NEGF+ncDFT approach reviewed in this Chapter can handle arbitrary device geometry, such as spin-valves and MTJs exhibiting STT or bilayers of the type FM/spin-orbit-coupled-material which are made inhomogeneous by attachment to NM leads, at vanishing or finite applied bias voltage. In contrast to often employed  2D transport geometry~\cite{Manchon2008,Haney2013,Lee2015,Li2015,Pesin2012a,Ado2017,Kalitsov2017,Ndiaye2017} for SOT theoretical analyses, we emphasize the importance of 3D transport geometry~\cite{Kim2017,Ghosh2018} to capture both the effects at the FM/spin-orbit-coupled-material interface and those further into the bulk of the FM layer. Finally, ultrathin FM layers employed in SOT experiments can hybridize strongly with the adjacent spin-orbit-coupled-material  to acquire its SOC and the corresponding spin textures on all of the FM 
monolayers. Such ``hybridized ferromagnetic metals'' can have electronic and spin structure (Fig.~\ref{fig:fig2}) which is quite different from an isolated FM layer, thereby requiring usage of both 3D geometry and first-principles Hamiltonians (of either tight-binding~\cite{Freimuth2014,Mahfouzi2018} or pseudopotential-LCAO-ncDFT~\cite{Theurich2001} type) to predict  the strength of SOT in realistic systems and optimal materials combinations for device applications of the SOT phenomenon.

\begin{acknowledgments}
We are grateful to K. D. Belashchenko, K. Xia and Z. Yuan for illuminating discussions and P.-H. Chang,  F. Mahfouzi and J.-M. Marmolejo-Tejada for collaboration. B. K. N. and K. D. were supported by DOE Grant No. DE-SC0016380 and NSF Grant No. ECCS 1509094. M.~P. and P.~P. were supported by ARO MURI Award No. W911NF-14-0247. K. S. and T. M.  acknowledge support from the European Commission Seventh Framework Programme Grant Agreement IIIV-MOS, Project No. 61932; and Horizon 2020 research and innovation programme under grant agreement SPICE, Project No. 713481. The supercomputing time was provided by XSEDE, which is supported by NSF Grant No.  ACI-1548562. 
\end{acknowledgments}

%BibTeX
%Windows:
%\bibliographystyle{C:/BIBTEX/IEEEtran}
%\bibliography{C:/BIBTEX/qttg}

\end{document}